\documentclass[sigconf]{acmart}
\usepackage{siunitx}
\usepackage{booktabs}
\usepackage{siunitx}
\usepackage{tabularx}
\usepackage{amsmath}
\usepackage{comment}
\usepackage{mathtools}

\AtBeginDocument{%
  }

\setcopyright{acmlicensed}
\copyrightyear{2018}
\acmYear{2026}
\acmDOI{XXXXXXX.XXXXXXX}
\acmConference[KDD '26]{32nd ACM SIGKDD Conference on Knowledge Discovery and Data Mining}{August 09--13,
  2026}{Jeju, Korea}
\acmISBN{978-1-4503-XXXX-X/2018/06}

\begin{document}

\title{A Data-Free, Physics-Informed Surrogate Solver for Drift Kinetic Equation: Enabling Fast Neoclassical Toroidal Viscosity Torque Modeling in Tokamaks}

\author{Xingting Yan}
\affiliation{%
  \institution{Hefei Institutes of Physical Science}
  \city{Hefei}
  \country{China}
}
\email{xingting.yan@ipp.ac.cn}

\author{Yuetao Meng$^{*}$}
\affiliation{%
  \institution{Anhui University}
  \city{Hefei}
  \country{China}}
\email{y23301016@stu.ahu.edu.cn}

\author{Nana Bao$^{*}$}
\affiliation{%
  \institution{Anhui University}
  \institution{Hefei Institutes of Physical Science}
  \city{Hefei}
  \country{China}
}
\email{nnbao@ahu.edu.cn}

\author{Youwen Sun}
\affiliation{%
  \institution{Hefei Institutes of Physical Science}
  \city{Hefei}
  \country{China}
}
\email{ywsun@ipp.ac.cn}

\author{Weiyong Zhou}
\affiliation{%
	\institution{Anhui University}
	\city{Hefei}
	\country{China}}
\email{y23301040@stu.ahu.edu.cn}

\author{Jinpeng Huang}
\affiliation{%
	\institution{Anhui University}
	\city{Hefei}
	\country{China}}
\email{y42314140@stu.ahu.edu.cn}



\begin{abstract}
Toroidal rotation is crucial for maintaining stable and high performance plasmas in tokamak fusion reactors. Among its driving mechanisms, the neoclassical toroidal viscosity (NTV) torque--induced by three-dimensional magnetic perturbations--is particularly significant due to its strong impact and controllability, especially for reactor-scale devices like ITER where conventional momentum injection method becomes less effective. However, traditional first-principle NTV modeling is computationally expensive, as it requires solving the drift kinetic equation (DKE) in high-dimensional phase space, therefore precluding any real-time applications such as active control or nonlinear integrated modeling of tokamak plasma. Although surrogate solver shows promising ability for accelerating scientific computations, obtaining the data required to train such model is still very challenging. In this work, we present a novel, data-free approach for developing fast surrogate solver of DKE, by training neural network solely based on physical constraints. Such physical constraints are implemented in two ways: First, the loss function is defined based on physical governing equations; Second, the boundary condition is hard-coded into the predicting model. The proposed model is validated against the dataset generated by first-principle numerical solver, which is found to achieve accurate DKE solution with significantly reduced time consuming. In particular, physics-driven surrogate shows higher physical consistency than data-driven surrogate. In general, our study provides a new idea for developing surrogate solvers in data-scarce scenarios, and demonstrates the potential of purely physics-driven neural networks to accelerate demanding scientific computations.
\end{abstract}

\ccsdesc[500]{Computing methodologies~Neural networks}
\ccsdesc[500]{Computing methodologies~Model verification and validation}
\ccsdesc[500]{Applied computing~Physics}

\keywords{Physics-Informed Machine Learning, Data-Free Training, Surrogate Solver, Drift Kinetic Equation, Neoclassical Toroidal Viscosity, Tokamak Physics}


\maketitle

\section{Introduction}

The toroidal rotation of magnetically-confined plasma significantly impacts tokamak performance, particularly regarding magnetohydrodynamic (MHD) instability, transport, and confinement~\cite{BecouletM_2017_NF,WuX_2025_NF}. Among various rotation drivers, the neoclassical toroidal viscosity (NTV) torque~\cite{Shaing_2003_POP,Shaing_2015_NF,Park_2009_PRL}—induced by three-dimensional (3D) magnetic perturbations—stands out for two reasons. First, it strongly influences toroidal momentum transport~\cite{Yang_2019_PRL,LoganN_2022_PRL}. Second, it can be actively controlled through external coils~\cite{KimS_2024_NC,LoganN_2021_NF}. In future reactor-scale devices like ITER (International Thermonuclear Experimental Reactor)~\cite{LoarteA_2025_NF}, conventional rotation control methods (e.g., neutral beam injection) become less effective due to large plasma inertia, making NTV torque a critical mechanism for active rotation control. A thorough understanding of NTV physics and related plasma dynamics is therefore essential, typically relying on first-principle, large-scale integrated numerical modelings.

However, traditional first-principle NTV modeling is computationally expensive, as it requires solving the governing drift kinetic equation (DKE) over typically tens of thousands of numerically discrete points in a high-dimensional phase space~\cite{Sun_2010_PRL,Sun_2019_POP}. This cost precludes any real-time applications, for example active control and self-consistent nonlinear integrated modeling of tokamak plasma. While deep learning offers a promising path toward surrogate modeling that balances physical accuracy with computational efficiency~\cite{Clement_2021_NF,Yan_2025_CPC}, such approaches usually depend on large, high-quality datasets, which are often scarce or unavailable.

Inspired by physics-informed neural network (PINN)~\cite{RaissiM_2019_JCP,CuomoS_2022_JSC,KarniadakisG_2021_NRP,ChoW_2024_ICML}, this work proposes a data-free, physics-constrained method for developing a surrogate DKE solver by directly embedding physical laws into the learning objective. The main contributions are:

\begin{itemize}
	\item A neural network surrogate model trained solely based on physical constraints (i.e. physical loss function and hard-coded boundary condition) rather than labeled data, enabling model training under data-scarce or data-free conditions.
	\item Improved physical consistency in both single-sample predictions and overall physical loss, compared to conventional data-driven surrogates.
	\item A computationally efficient model capable of predicting DKE solutions across varying equation parameters, thus enabling fast NTV calculations.
\end{itemize}

\begin{figure}[htbp]
	\centering
	\includegraphics[width=0.48\textwidth]{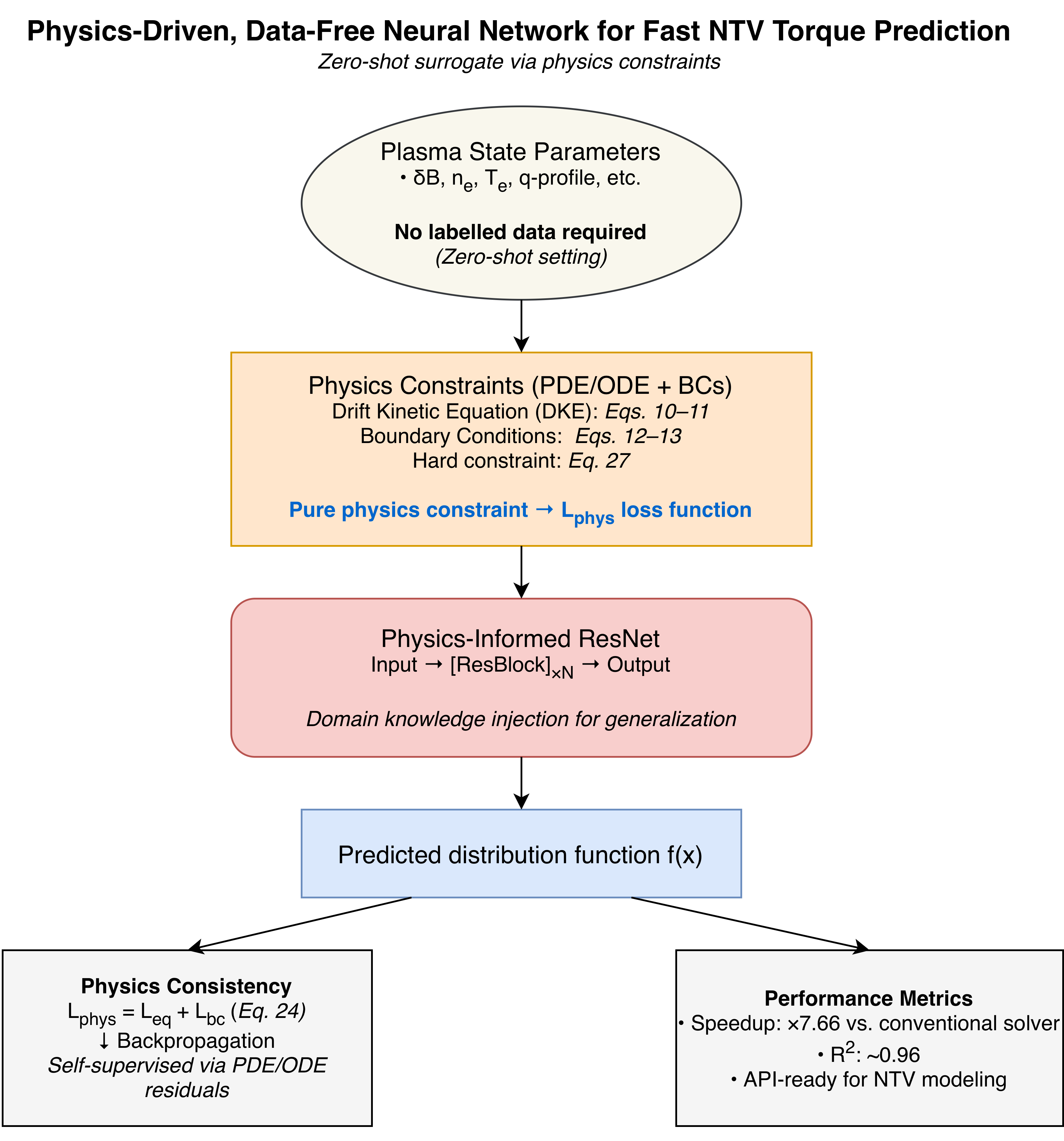}
	\caption{Key ideas and results of this paper.}
	\label{fig:flowchart}
\end{figure}

The key ideas of this work are illustrated by Fig.~\ref{fig:flowchart}. The remainder of this paper is organized as follows: The physics background and problem definition (i.e. DKE and NTV) are illustrated in Sec.~\ref{sec:ntv_dke}. The machine learning method adopted in this work, including the model, loss function and the hard-coded boundary condition are introduced in Sec.~\ref{sec:methods}. The main results are shown in Sec.~\ref{sec:res}, where various aspects of model performance are compared and discussed. Finally, summary and future work are given in Sec.~\ref{sec:sum}.

\section{Drift kinetic equation and NTV torque}
\label{sec:ntv_dke}

The drift kinetic equation in NTV torque modeling is a complex-valued ordinary differential equation, defined on the pitch angle variant coordinate~\cite{Sun_2019_POP}:
\begin{equation}
	\kappa^2\triangleq\frac{v^2/2-\mu B_m}{\mu(B_M-B_m)}\in[0,1],
\end{equation}
where $v$ is the particle velocity, $\mu$ is the magnetic moment, $B_M$ and $B_m$ are the maximum and minimum magnetic field strength on one flux surface respectively. The DKE can be written as:
\begin{equation}\label{eq:dke}
	c_1F_2d_{\kappa^2}^2f+c_1F_1d_{\kappa^2}f+c_2f-c_3=0,
\end{equation}
where $c_1$, $c_2$, $c_3$, $F_1$, $F_2$ are all DKE parameters defined on $\kappa^2$ coordinate:
\begin{equation}
	c_1=\frac{\nu_d/(2\epsilon)}{I_0},
\end{equation}
\begin{equation}
	c_2=i\frac{\Omega_{nl}}{I_0}+c_1F_2(d_{\kappa^2}D)^2,
\end{equation}
\begin{equation}
	F_1=\frac{1}{p}(1+4\epsilon\frac{k_b^2}{1+2\epsilon B_\epsilon}),
\end{equation}
\begin{equation}
	F_2=\frac{2}{p^2}\frac{k_b^2}{1+2\epsilon B_\epsilon},
\end{equation}
\begin{equation}
	D=l\eta+n\Delta,
\end{equation}
$c_3$ represents the Fourier harmonic of magnetic perturbation, $\nu_d$ is the deflection frequency, $\epsilon$ is the inverse aspect ratio, $I_0$ is a normalization parameter, $\Omega_{nl}$ is the combination of precession and bounce frequencies, $p=1/(1+2\epsilon\kappa^2)$, $k_b=\sqrt{\kappa^2-B_\epsilon}$, $B_\epsilon=(B-B_m)/(B_M-B_m)$, $B$ is magnetic field strength on one flux surface, $n, l$ are Fourier mode numbers, $\eta$ and $\Delta$ are coordinate-related parameters. $f$ is the solution to DKE, i.e. the Fourier harmonic of plasma perturbed distribution function. To uniquely determine the equation solution, two boundary conditions are required~\cite{Sun_2011_NF}:
\begin{equation}\label{eq:bc}
	\begin{aligned}
		F_2d_{\kappa^2}f|_{\kappa^2=0}&=0, \\
		f|_{\kappa^2=1}&=0.
	\end{aligned}
\end{equation}
In one complete NTV calculation cycle, Eq.~\ref{eq:dke} is solved tens of thousands times under various DKE parameters corresponding to different discrete points in high-dimensional phase space, which leads to the main computational bottleneck. The NTV torque ($T_{NTV}$) can then be calculated through straightforward phase space integral calculation of $f$. 

To be compatible with machine learning framework, Eq.~\ref{eq:dke} defined in complex space needs to be transformed to real space. For this purpose, the complex-valued functions in Eq.~\ref{eq:dke} ($c_2$, $c_3$ and $f$) are separated as real and imaginary parts:
\begin{equation}\label{eq:coef_ri}
	\begin{aligned}
		c_2&=c_{2r}+ic_{2i}, \\
		c_3&=c_{3r}+ic_{3i}, \\
		f&=f_r+if_i,
	\end{aligned}
\end{equation}
where the subscripts '$r$' and '$i$' indicate real and imaginary parts. By substituting Eq.~\ref{eq:coef_ri} into Eq.~\ref{eq:dke} and Eq.~\ref{eq:bc}, we obtain the real and imaginary parts of DKE:
\begin{equation}\label{eq:eq_r}
	c_1F_2d_{\kappa^2}^2f_r+c_1F_1d_{\kappa^2}f_r+c_{2r}f_r-c_{2i}f_i-c_{3r}=0,
\end{equation}
\begin{equation}\label{eq:eq_i}
	c_1F_2d_{\kappa^2}^2f_i+c_1F_1d_{\kappa^2}f_i+c_{2r}f_i+c_{2i}f_r-c_{3i}=0,
\end{equation}
the real parts of boundary conditions:
\begin{equation}\label{eq:bc_r}
	\begin{aligned}
		F_2d_{\kappa^2}f_r|_{\kappa^2=0}&=0, \\
		f_r|_{\kappa^2=1}&=0,
	\end{aligned}
\end{equation}
and the imaginary parts:
\begin{equation}\label{eq:bc_i}
	\begin{aligned}
		F_2d_{\kappa^2}f_i|_{\kappa^2=0}&=0, \\
		f_i|_{\kappa^2=1}&=0.
	\end{aligned}
\end{equation}
The objective of surrogate solver is to predict the DKE solution ($f_r$, $f_i$) for any given set of DKE parameters ($F_1$, $F_2$, $c_1$, $c_{2r}$, $c_{2i}$, $c_{3r}$, $c_{3i}$) constrained by Eqs.~\ref{eq:eq_r}-\ref{eq:bc_i}.

\section{Methods}
\label{sec:methods}

\subsection{Dataset}

The dataset is constructed by generating multiple instances of mapping from DKE parameters to DKE solution. In this paper, DKE parameters are generated based on experimental conditions of the Experimental Advanced Superconducting Tokamak (EAST)~\cite{WanB_2017_NF}, the DKE solutions are obtained by traditional numerical differential equation solver implemented in NTVTOK (NTV in TOKmaks) code~\cite{Sun_2019_POP}, which utilizes finite difference (FD) method and LU decomposition. The $\kappa^2$ coordinate space is uniformly divided into $N_g$ numerical grid points. The DKE parameters are concatenated to construct the input feature for each data sample, i.e.
\begin{equation}
	{\bf{X}_{in}}=[F_1,F_2,c_1,c_{2r},c_{2i},c_{3r},c_{3i}],
\end{equation}
and the real and imaginary parts of DKE solution are concatenated to construct the output label for each data sample, i.e.
\begin{equation}
	{\bf{X}_{out}}=[f_r,f_i].
\end{equation}
In this work, there are $N_s=20000$ data samples (i.e. ${\bf{X}_{in}}$-${\bf{X}_{out}}$ pairs) in total, and the number of grid points $N_g=200$. For each data sample, $c_1$ is a constant, other DKE parameters and DKE solutions are functions of $\kappa^2$, thus ${\bf{X}_{in}}$ is a vector of length $6\cdot N_g+1$, ${\bf{X}_{out}}$ is a vector of length $2\cdot N_g$. The typical statistical features of each physical variable are shown in Table~\ref{tab:stats}.

\begin{table}[htbp]
	\centering
	\begin{tabular}{|c|c|c|c|c|}
		\hline
		\textbf{Variable} & {\textbf{Min}} & {\textbf{Max}} & {\textbf{Mean}} & {\textbf{Median}} \\ \hline
		$F_1$ & 1.00e+00 & 1.93e+00 & 1.21e+00 & 1.15e+00 \\
		$F_2$ & 9.90e-04 & 1.03e+00 & 4.91e-01 & 5.17e-01 \\
		$c_1$ & 1.71e-04 & 1.00e+00 & 4.26e-01 & 1.91e-01 \\
		$c_{2r}$ & -1.26e+06 & -8.37e-13 & -3.14e+01 & -2.82e-05 \\
		$c_{2i}$ & -1.00e+00 & 1.00e+00 & 3.46e-01 & 3.44e-01 \\
		$c_{3r}$ & -1.00e+00 & 1.00e+00 & -9.50e-02 & -6.23e-02 \\
		$c_{3i}$ & -1.34e-01 & 9.55e-01 & 1.36e-02 & -1.34e-03 \\ \hline
		$f_r$ & -5.24e+00 & 1.45e+01 & 1.13e-02 & 7.68e-03 \\
		$f_i$ & -1.45e+01 & 4.12e+00 & 4.94e-02 & 1.97e-03 \\ \hline
	\end{tabular}
	\caption{Statistics of Input and Output Variables.}
	\label{tab:stats}
\end{table}

\subsection{Loss function}

The left hand sides (LHS) of Eqs.~\ref{eq:eq_r}-\ref{eq:eq_i} can be evaluated on each numerical grid point in $\kappa^2$ coordinate space for each data sample. Let:
\begin{equation}
	LHS_r=c_1F_2d_{\kappa^2}^2f_r+c_1F_1d_{\kappa^2}f_r+c_{2r}f_r-c_{2i}f_i-c_{3r},
\end{equation}
\begin{equation}
	LHS_i=c_1F_2d_{\kappa^2}^2f_i+c_1F_1d_{\kappa^2}f_i+c_{2r}f_i+c_{2i}f_r-c_{3i},
\end{equation}
then we define:
\begin{equation}
	\mathcal{L}_{eq,r}^{ig,is}\coloneqq (LHS_r)^2|_{ig\in[1:N_g],is\in[1:N_s]}
\end{equation}
and
\begin{equation}
	\mathcal{L}_{eq,i}^{ig,is}\coloneqq (LHS_i)^2|_{ig\in[1:N_g],is\in[1:N_s]}
\end{equation}
For boundary conditions in Eqs.~\ref{eq:bc_r}-\ref{eq:bc_i}, the following terms can be defined in similar way:
\begin{align}
	&\mathcal{L}_{bc,r0}^{is}\coloneqq (F_2d_{\kappa^2}f_r|_{\kappa^2=0})^2|_{is\in[1:N_s]}, \\
	&\mathcal{L}_{bc,r1}^{is}\coloneqq (f_r|_{\kappa^2=1})^2|_{is\in[1:N_s]}, \\
	&\mathcal{L}_{bc,i0}^{is}\coloneqq (F_2d_{\kappa^2}f_i|_{\kappa^2=0})^2|_{is\in[1:N_s]}, \\
	&\mathcal{L}_{bc,i1}^{is}\coloneqq (f_i|_{\kappa^2=1})^2|_{is\in[1:N_s]}.
\end{align}
The total physics-constrained loss function is written as:
\begin{equation}
	\begin{aligned}
		\mathcal{L}_{phys}&=\lambda_{eq,r}\cdot\mathcal{L}_{eq,r}+ \lambda_{eq,i}\cdot\mathcal{L}_{eq,i} \\
		&+\lambda_{bc,r0}\cdot\mathcal{L}_{bc,r0}+\lambda_{bc,i0}\cdot\mathcal{L}_{bc,i0} \\
		&+\lambda_{bc,r1}\cdot\mathcal{L}_{bc,r1}+\lambda_{bc,i1}\cdot\mathcal{L}_{bc,i1},
	\end{aligned}
\end{equation}
where
\begin{equation}
	\begin{aligned}
		&\mathcal{L}_{eq,r}=\left\langle\mathcal{L}_{eq,r}^{ig,is}\right\rangle_{ig,is}, \mathcal{L}_{eq,i}=\left\langle\mathcal{L}_{eq,i}^{ig,is}\right\rangle_{ig,is}, \\
		&\mathcal{L}_{bc,r0}=\left\langle\mathcal{L}_{bc,r0}^{is}\right\rangle_{is}, \mathcal{L}_{bc,i0}=\left\langle\mathcal{L}_{bc,i0}^{is}\right\rangle_{is}, \\
		&\mathcal{L}_{bc,r1}=\left\langle\mathcal{L}_{bc,r1}^{is}\right\rangle_{is}, \mathcal{L}_{bc,i1}=\left\langle\mathcal{L}_{bc,i1}^{is}\right\rangle_{is},
	\end{aligned}
\end{equation}
where the superscripts '$ig$' denotes grid point index, '$is$' denotes sample index, $\left\langle ... \right\rangle$ denotes average over all grid points and data samples, $\lambda_{eq,r}$, $\lambda_{eq,i}$, $\lambda_{bc,r0}$, $\lambda_{bc,i0}$, $\lambda_{bc,r1}$, $\lambda_{bc,i1}$ are adjustable weighting parameters.

A data-driven surrogate is also trained in this work for model comparison. The corresponding learning objective is the conventional MSE (mean squared error) loss function:
\begin{equation}
	\mathcal{L}_{data}=\frac{1}{N_s\cdot 2N_g}\sum_{ig,is}\left[\left(f_r^{ig,is}-\hat{f}_r^{ig,is}\right)^2+\left(f_i^{ig,is}-\hat{f}_i^{ig,is}\right)^2\right],
\end{equation}
where $f$ and $\hat{f}$ denote data label and prediction, respectively.

\subsection{Hard constraint at $\kappa^2=1$ boundary}

The boundary condition at $\kappa^2=1$ (i.e. $f|_{\kappa^2=1}=0$) can either be constrained by minimizing the loss function, or hard-coded in model structure. For the latter, this hard constraint can be realized by multiplying the model output directly with $1-\kappa^2$, i.e.
\begin{equation}
	Output_{NN}=Output_{NN}\cdot(1-\kappa^2).
\end{equation}
Consequently, $\mathcal{L}_{bc,r1}$ and $\mathcal{L}_{bc,i1}$ terms automatically become zero. The effectiveness of this strategy will be shown in the following section.

\section{Results and discussions}
\label{sec:res}

\subsection{Experiment setup}

Residual Neural Network (ResNet)~\cite{HeK_2016_CVPR} is adopted as the deep learning model in this work, the basic model parameters are shown in Table~\ref{tab:model_paras}. Three surrogates as listed in Table~\ref{tab:models} are compared in detail from various aspects, which differ in loss function definition and boundary constraint handling.

\begin{table}[htbp]
	\centering
	\begin{tabular}{|c|c|c|}
		\hline
		\bf{Parameters} & \bf{Meaning} & \bf{Value} \\
		\hline
		num\_blocks & Number of residual blocks & 2 \\
		activation & Activation function & $\tanh$ \\
		hidden\_size & Hidden layer dimension & 512 \\
		residual\_scale & Residual scaling factor & 0.5 \\
		epochs & Training cycle & 50000 \\
		\hline
	\end{tabular}
	\caption{Basic model and training parameters.}
	\label{tab:model_paras}
\end{table}

\begin{table}[htbp]
	\centering
	\begin{tabular}{|c|c|c|c|}
		\hline
		\bf{Name} & \bf{Loss} & \bf{Model} & \bf{Hard constraint} \\
		\hline
		$DKE_{data}$ & $\mathcal{L}_{data}$ & ResNet & No \\
		$DKE_{phys}$ & $\mathcal{L}_{phys}$ & ResNet & No \\
		$DKE_{phys,bc1}$ & $\mathcal{L}_{phys}$ & ResNet & Yes, ResNet$\times(1-\kappa^2)$ \\
		\hline
	\end{tabular}
	\caption{Difference of surrogate models.}
	\label{tab:models}
\end{table}

For simplicity, the weighting parameters in $\mathcal{L}_{phys}$ are all equal to $1$. The total $20000$ data samples are randomly split into training and validation sets as the ratio of $8:2$. All experiments are executed on the same Linux workstation equipped with Intel Xeon Gold 6248 CPU @ 2.50GHz, 128GB RAM, and NVIDIA GeForce RTX 4090 GPU.

\subsection{Overall model performance}

\begin{figure}[htbp]
	\centering
	\includegraphics[width=0.22\textwidth]{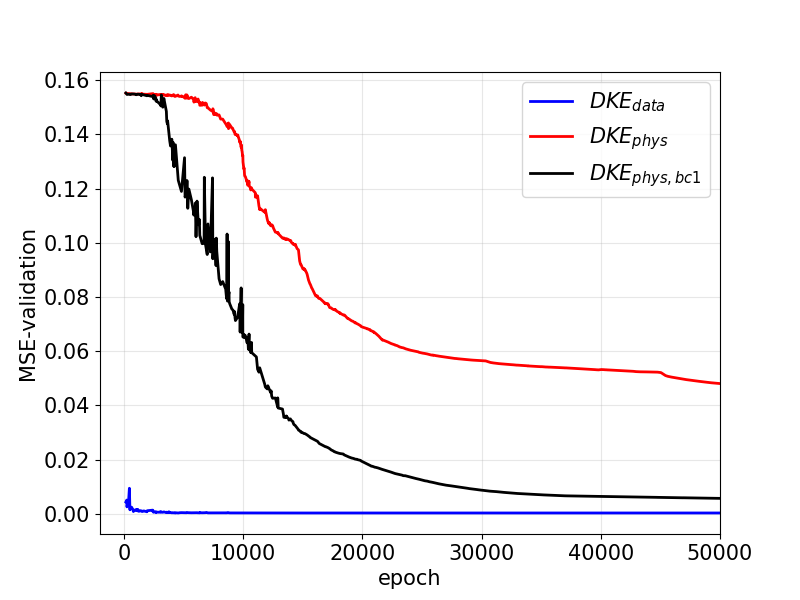}
	\put(-20,40){\textbf{\footnotesize{(a)}}}
	\includegraphics[width=0.22\textwidth]{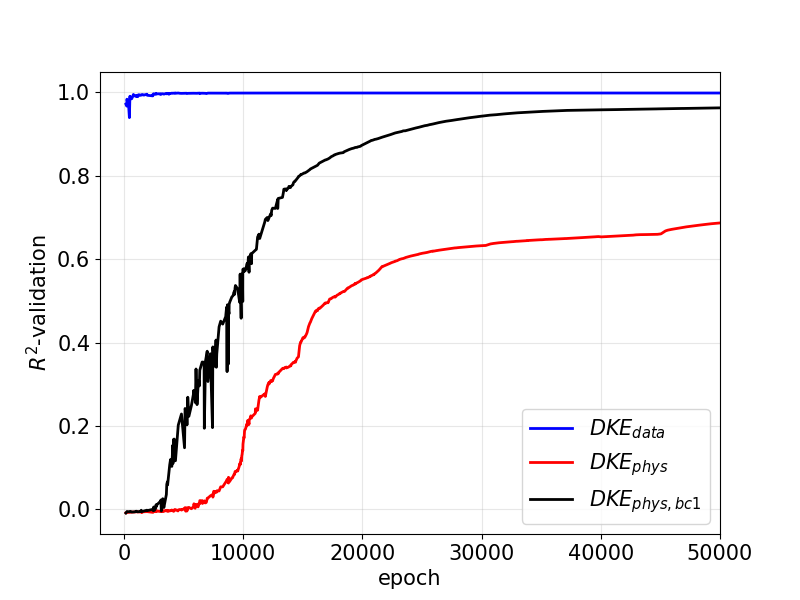}
	\put(-20,40){\textbf{\footnotesize{(b)}}}
	\caption{Evolution of (a) MSE and (b) $R^2$ on validation set during model training for three surrogates: $DKE_{data}$, $DKE_{phys}$ and $DKE_{phys,bc1}$.}
	\label{fig:data_loss}
\end{figure}

During the training processes for all three surrogates, various evaluation metrics and loss terms are evaluated on the validation set to monitor the model performance. Fig.~\ref{fig:data_loss} shows evolution of (a) MSE loss and (b) $R^2$ for three surrogates, representing the data proximity between ground truth and prediction. It is obvious that $DKE_{data}$ (blue line) converges to very high accuracy regarding MSE and $R^2$ very quickly, because its learning objective (i.e. MSE loss) is a direct measure of data proximity. The surrogate $DKE_{phys}$, purely trained by minimizing physics loss, shows relatively poor performance. However, the surrogate $DKE_{phys,bc1}$ which incorporates both physics loss and hard-constrained boundary condition exhibits similar performance with $DKE_{data}$, indicating both the feasibility of physics-driven training approach and the important role of boundary condition handling.

In addition to the standard evaluation metrics like MSE and $R^2$, the relative prediction error $\varepsilon_s$ is also defined to measure model accuracy for each single data sample:
\begin{equation}
	\varepsilon_s=\frac{|f-\hat{f}|}{\max(|f|)},
\end{equation}
where $f$ and $\hat{f}$ are ground truth and model prediction respectively, '$\max$' operation is performed over the $N_g$ grid points for each data sample, the subscript '$s$' indicates single sample. The mean and median values of $\varepsilon_s$ over the whole dataset can also be used to evaluate model performance, which is shown in Table~\ref{tab:metrics} along with MSE and $R^2$. The best prediction accuracy regarding relative error is also achieved by $DKE_{data}$. Except $DKE_{phys}$ where the mean relative error is larger than $20\%$, all other models exhibit reasonable accuracy with mean and median relative error less than several percents.

\begin{table}[htbp]
	\centering
	\begin{tabular}{|c|c|c|c|c|}
		\hline
		\bf{Model} & \bf{MSE} & \bf{$R^2$} & \bf{$mean(\varepsilon_s)$} & \bf{$median(\varepsilon_s)$} \\
		\hline
		$DKE_{data}$ & 2.73e-04 & 0.99 & 5.89e-03 & 1.09e-03 \\
		$DKE_{phys}$ & 4.80e-02 & 0.69 & 2.87e-01 & 7.91e-02 \\
		$DKE_{phys,bc1}$ & 5.62e-03 & 0.96 & 3.44e-02 & 8.32e-03 \\
		\hline
	\end{tabular}
	\caption{Evaluation metrics of surrogate models.}
	\label{tab:metrics}
\end{table}

\begin{figure}[htbp]
	\centering
	\includegraphics[width=0.22\textwidth]{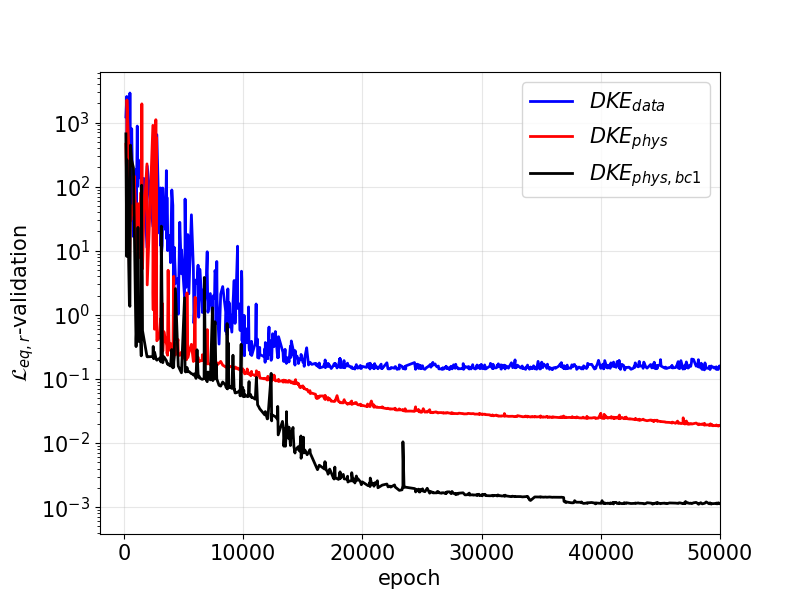}
	\put(-20,40){\textbf{\footnotesize{(a)}}}
	\includegraphics[width=0.22\textwidth]{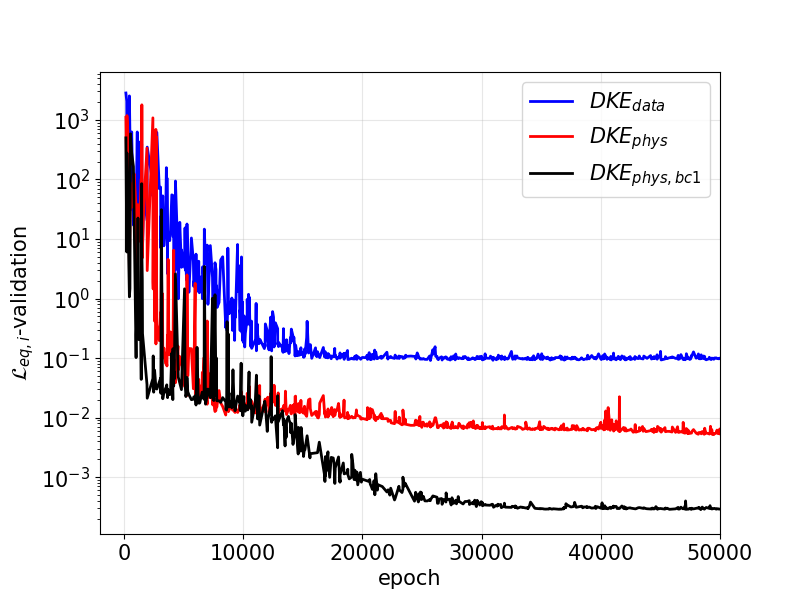}
	\put(-20,40){\textbf{\footnotesize{(b)}}}
	\vfill
	\includegraphics[width=0.22\textwidth]{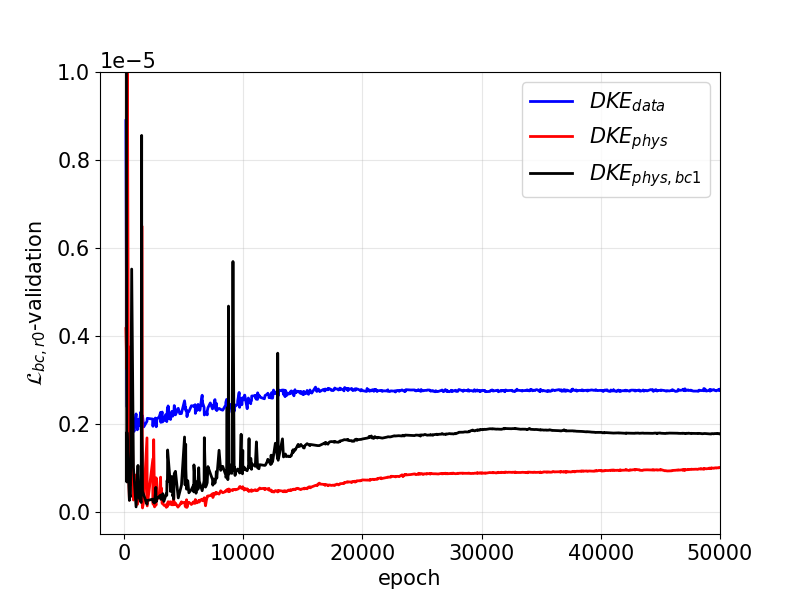}
	\put(-20,40){\textbf{\footnotesize{(c)}}}
	\includegraphics[width=0.22\textwidth]{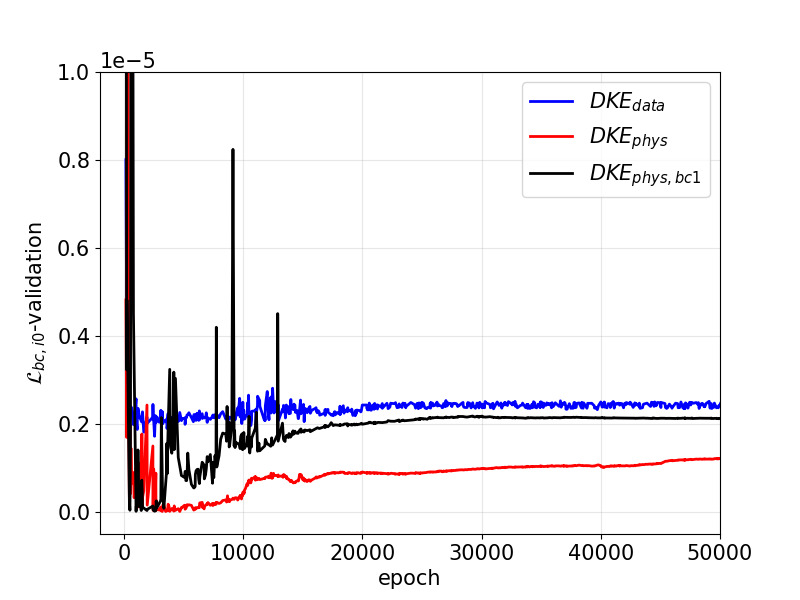}
	\put(-20,40){\textbf{\footnotesize{(d)}}}
	\vfill
	\includegraphics[width=0.22\textwidth]{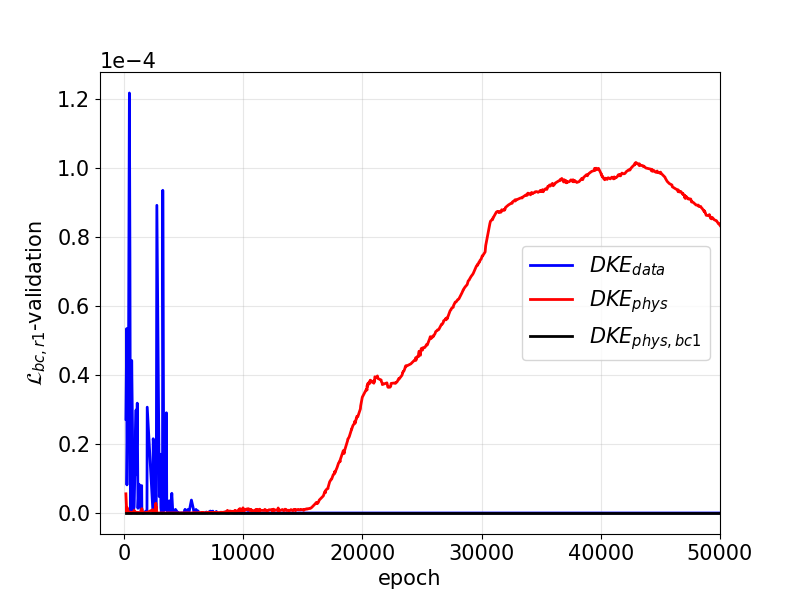}
	\put(-20,20){\textbf{\footnotesize{(e)}}}
	\includegraphics[width=0.22\textwidth]{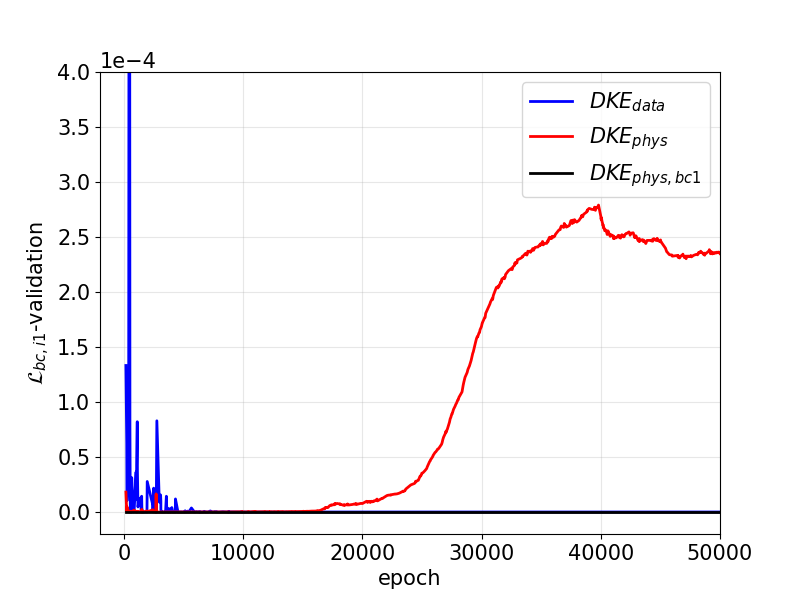}
	\put(-20,20){\textbf{\footnotesize{(f)}}}
	\caption{Evolution of different components of physical loss during model training: (a) $\mathcal{L}_{eq,r}$; (b) $\mathcal{L}_{eq,i}$; (c) $\mathcal{L}_{bc,r0}$; (d) $\mathcal{L}_{bc,i0}$; (e) $\mathcal{L}_{bc,r1}$; (f) $\mathcal{L}_{bc,i1}$.}
	\label{fig:phys_loss}
\end{figure}

Fig.~\ref{fig:phys_loss} shows the evolution of various physical loss terms during training process, including the equation loss, boundary condition loss at $\kappa^2=0$, boundary condition loss at $\kappa^2=1$, and the real/imaginary parts of them respectively. From Figs.~\ref{fig:phys_loss}(a) and (b), it shows that the lowest level of equation loss ($\mathcal{L}_{eq,r}$ and $\mathcal{L}_{eq,i}$) is achieved by $DKE_{phys,bc1}$ (black lines), which incorporates both physics loss function and hard-constrained boundary condition. The data-driven model--$DKE_{data}$, shows significantly larger equation loss (blue lines), indicating possible 'unphysical' predictions although the data proximity is very high. Similar phenomenon is observed in Figs.~\ref{fig:phys_loss}(c) and (d), where $DKE_{data}$ shows larger boundary condition loss at $\kappa^2=0$ than physics-driven models. For boundary condition loss at $\kappa^2=1$ in Figs.~\ref{fig:phys_loss}(e) and (f), since the hard-constraint is implemented in $DKE_{phys,bc1}$, its corresponding loss term is exactly zero. $DKE_{data}$ also shows good performance in reducing boundary condition loss, but $DKE_{phys}$ shows significant increase of $\mathcal{L}_{bc,r1}$ and $\mathcal{L}_{bc,r1}$ during training process (red lines), indicating the importance of hard-constrained boundary condition implemented in $DKE_{phys,bc1}$.

\subsection{Analysis on single sample prediction}

\begin{figure}[htbp]
	\centering
	\includegraphics[width=0.22\textwidth]{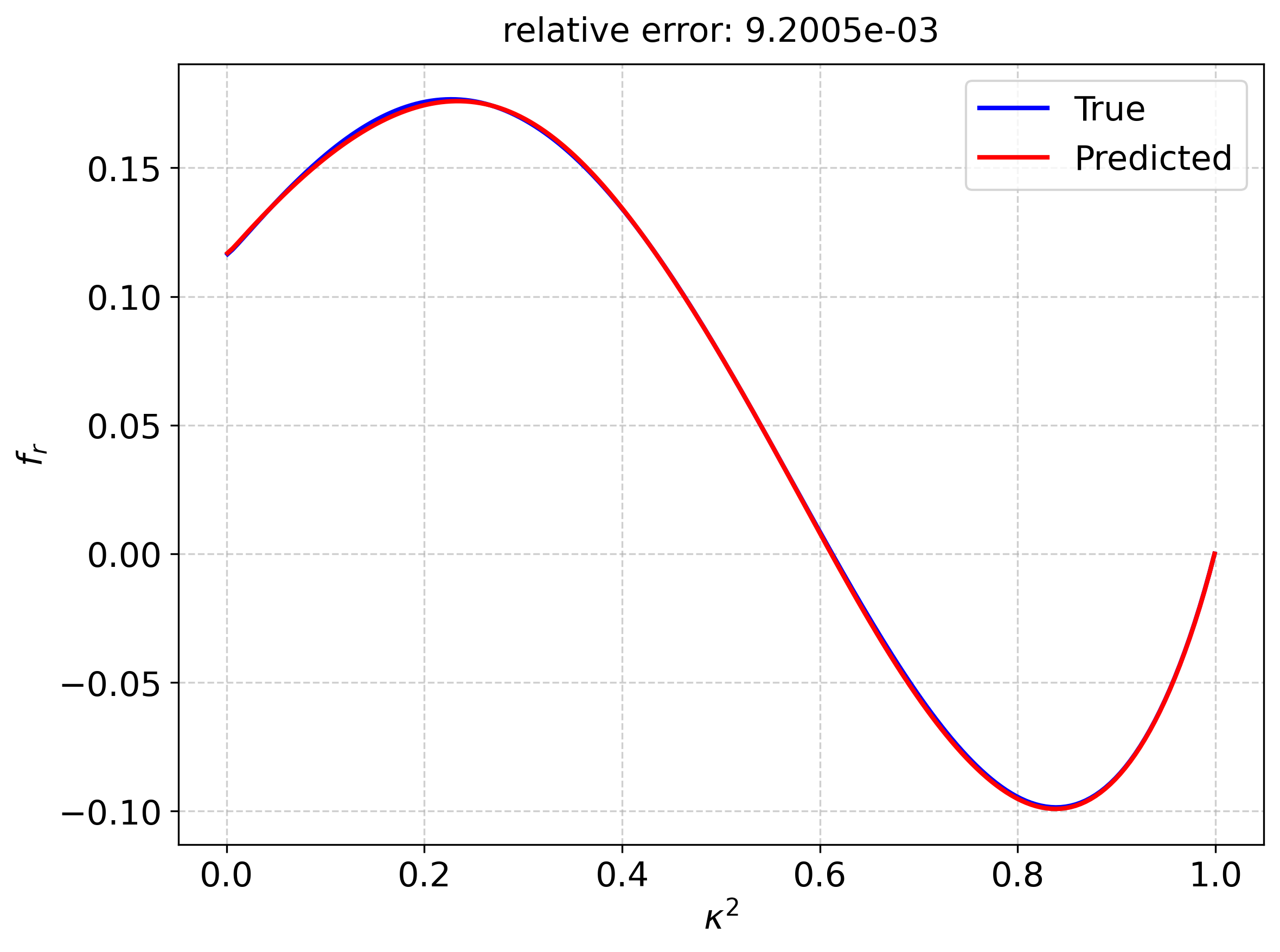}
	\put(-90,20){\textbf{\footnotesize{(a) $DKE_{data}$}}}
	\includegraphics[width=0.22\textwidth]{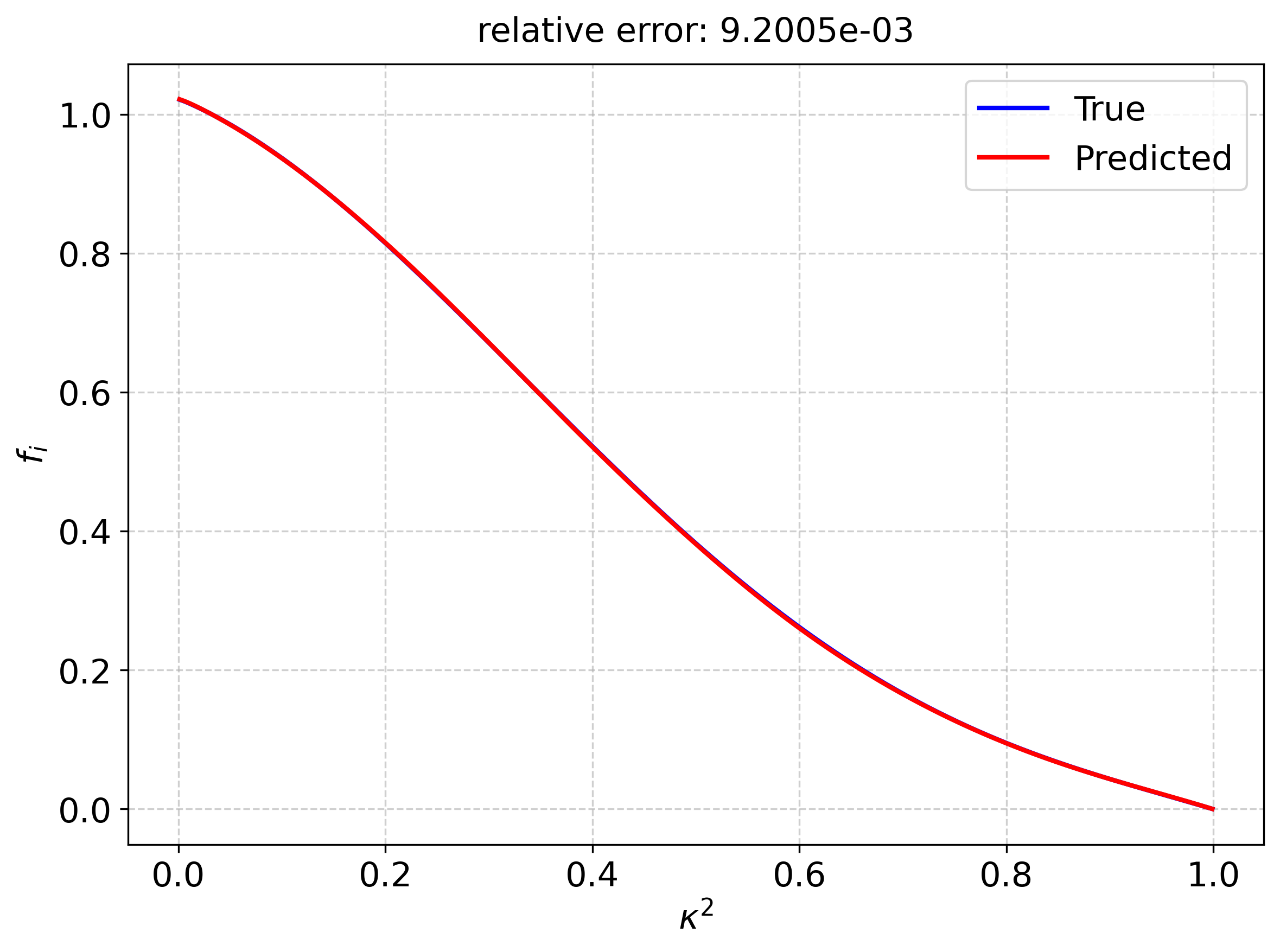}
	\put(-90,20){\textbf{\footnotesize{(b) $DKE_{data}$}}}
	\vfill
	\includegraphics[width=0.22\textwidth]{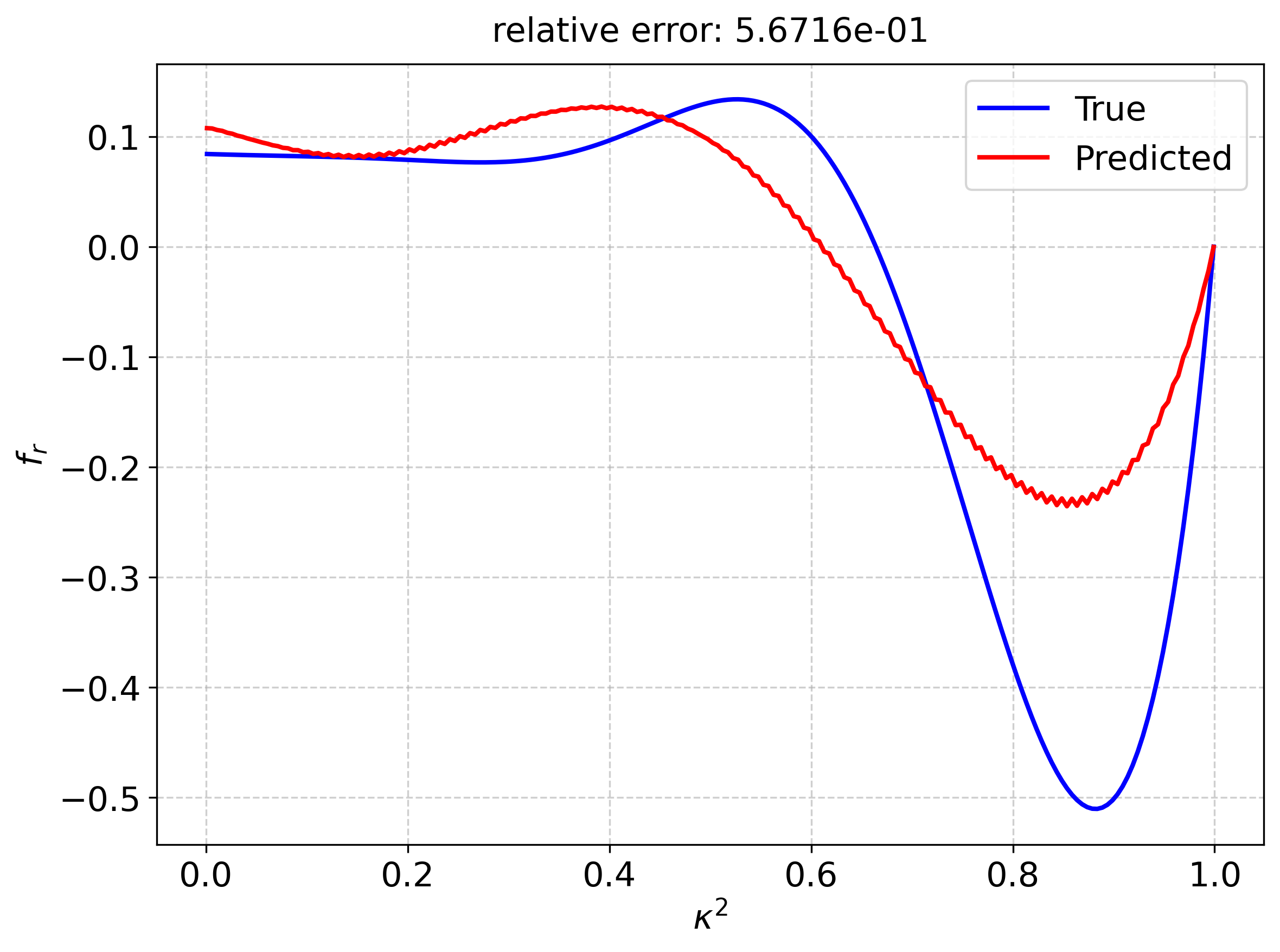}
	\put(-90,20){\textbf{\footnotesize{(c) $DKE_{phys}$}}}
	\includegraphics[width=0.22\textwidth]{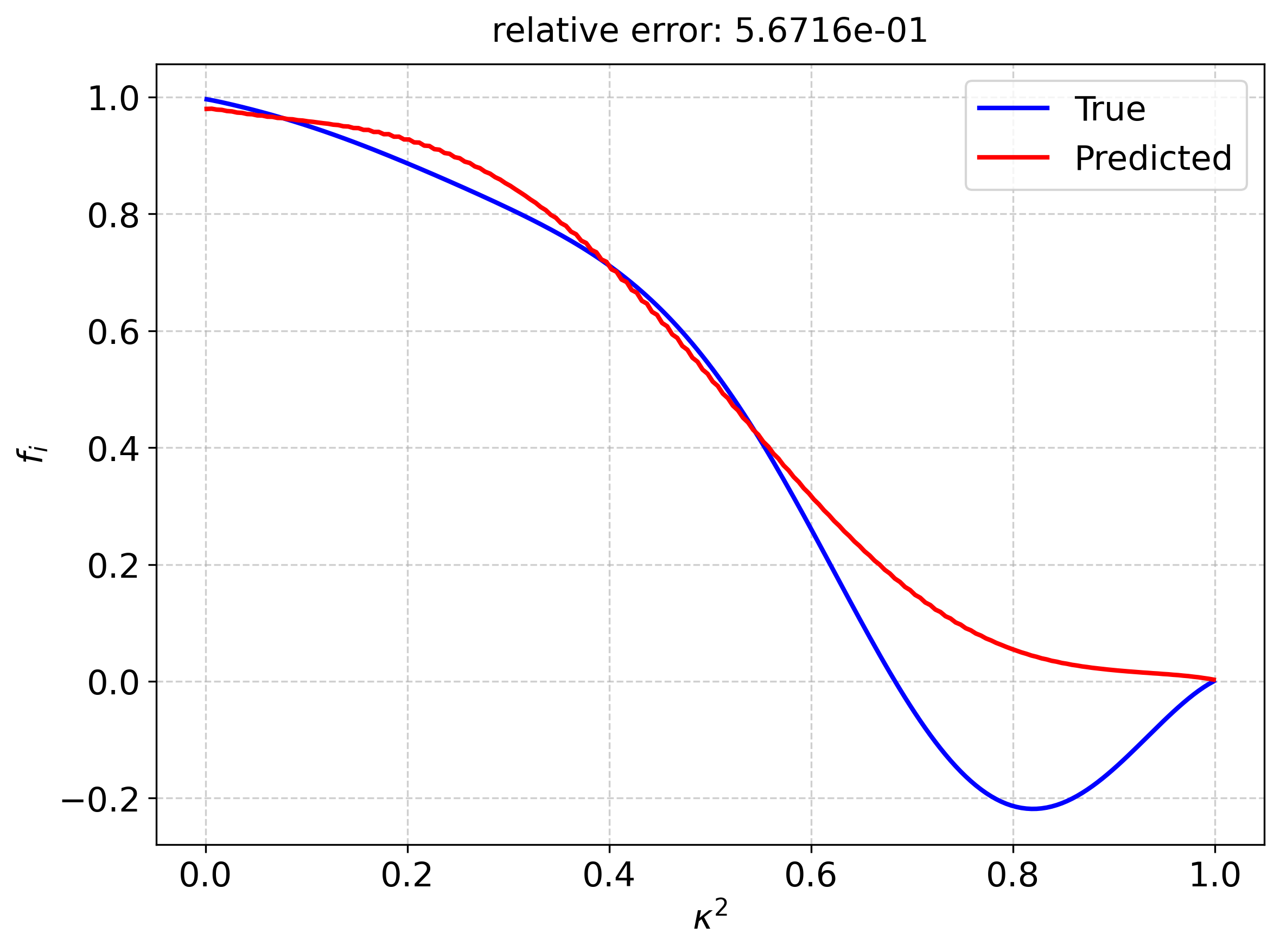}
	\put(-90,20){\textbf{\footnotesize{(d) $DKE_{phys}$}}}
	\vfill
	\includegraphics[width=0.22\textwidth]{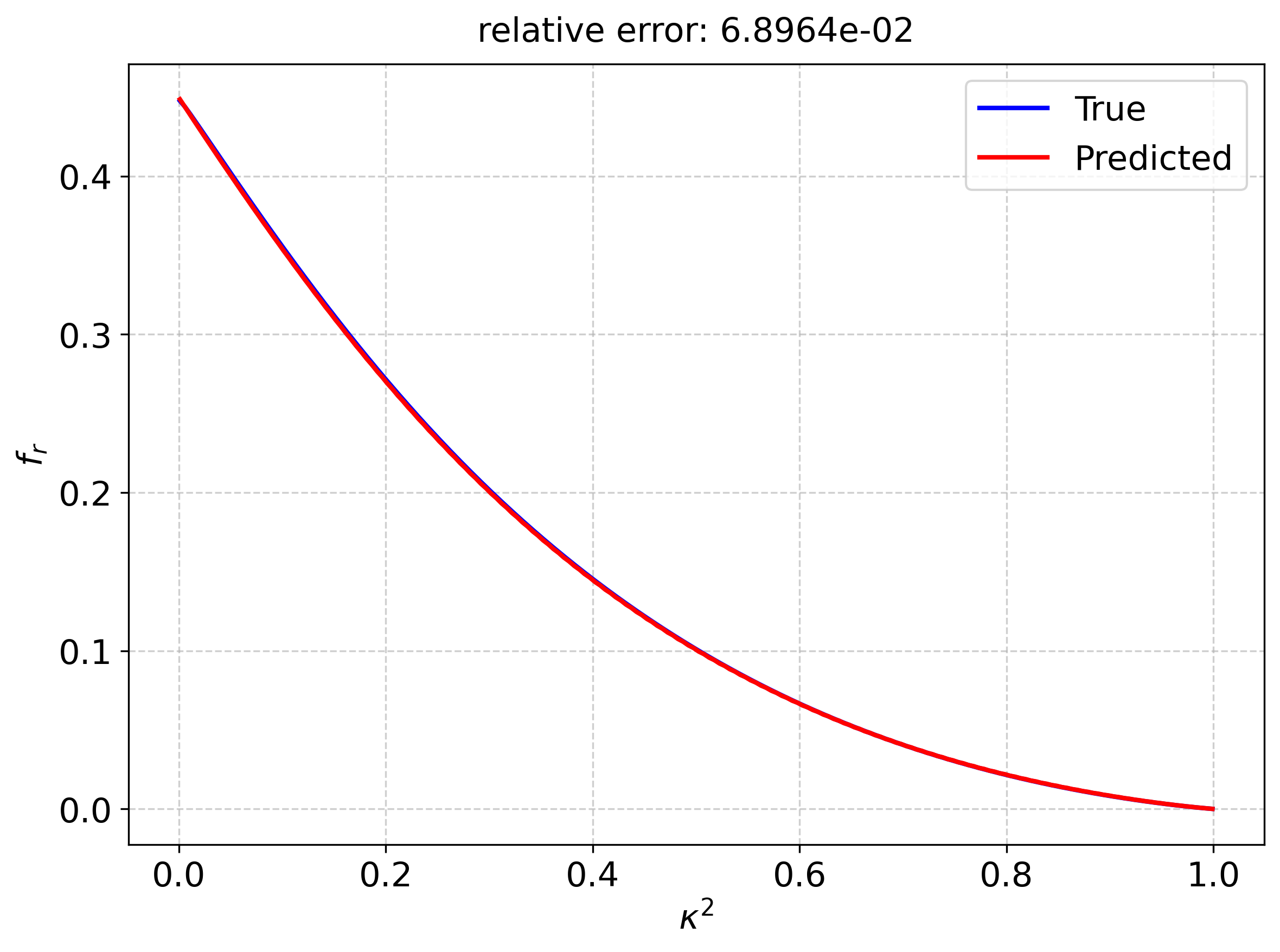}
	\put(-90,20){\textbf{\footnotesize{(e) $DKE_{phys,bc1}$}}}
	\includegraphics[width=0.22\textwidth]{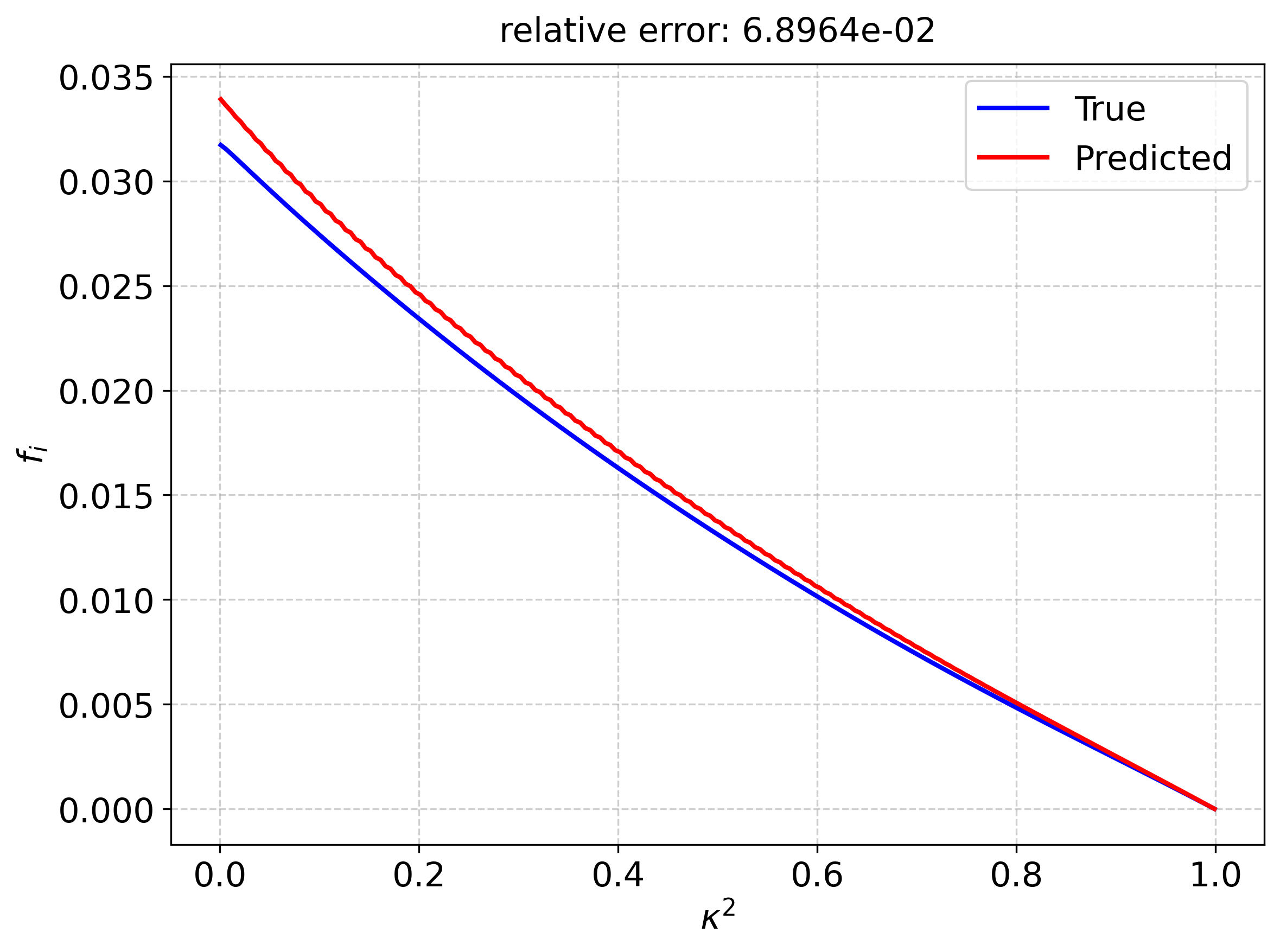}
	\put(-90,20){\textbf{\footnotesize{(f) $DKE_{phys,bc1}$}}}
	\caption{Prediction of DKE solution (left column for $f_r$ and right column for $f_i$) by surrogate models (top panel: $DKE_{data}$; middle panel: $DKE_{phys}$; bottom panel: $DKE_{phys,bc1}$) for data sample corresponding to median relative error.}
	\label{fig:median_sample}
\end{figure}

Fig.~\ref{fig:median_sample} shows the predicted profile of DKE solution (i.e. the perturbed distribution function) for data sample corresponding to median relative error, for $DKE_{data}$, $DKE_{phys}$ and $DKE_{phys,bc1}$ from top to bottom panels respectively. The left column corresponds to the real part $f_r$ and the right column for the imaginary part $f_i$. It shows that $DKE_{data}$ exactly reproduces the ground truth value for this specific data sample, consistent with its high data proximity. $DKE_{phys}$, however, shows poorer consistency with the ground truth, although the general profile shape is decently captured. The $DKE_{phys,bc1}$ surrogate shows high prediction accuracy as well, except for deviation around $\kappa^2\sim0$ region in imaginary part prediction. The performance of $DKE_{phys,bc1}$ can be further improved (currently $R^2\sim0.96$), by fine-tuning of hyper-parameters for example, which will remain for future work.

\begin{figure}[htbp]
	\centering
	\includegraphics[width=0.22\textwidth]{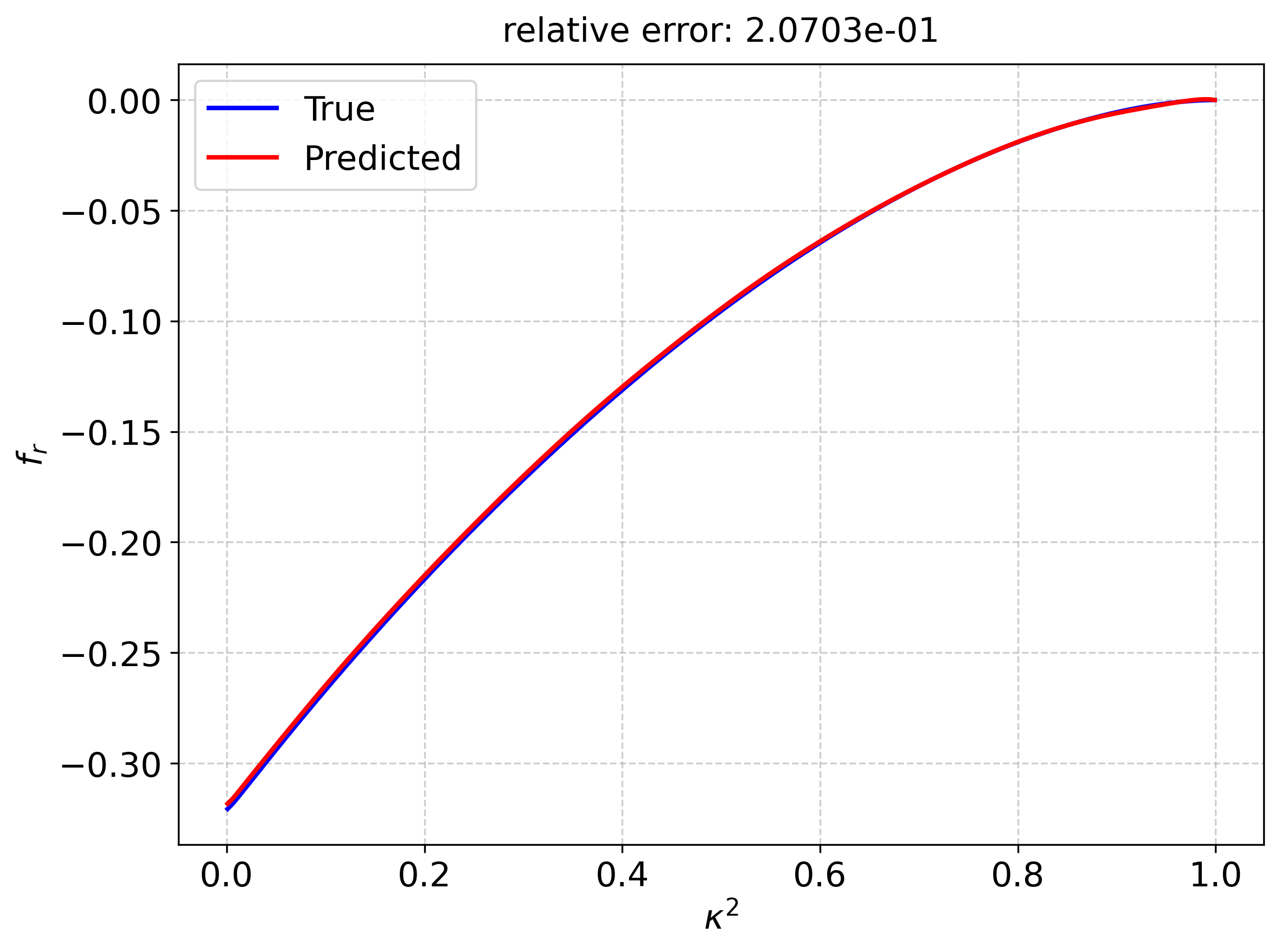}
	\put(-50,40){\textbf{\footnotesize{(a) $DKE_{data}$}}}
	\includegraphics[width=0.22\textwidth]{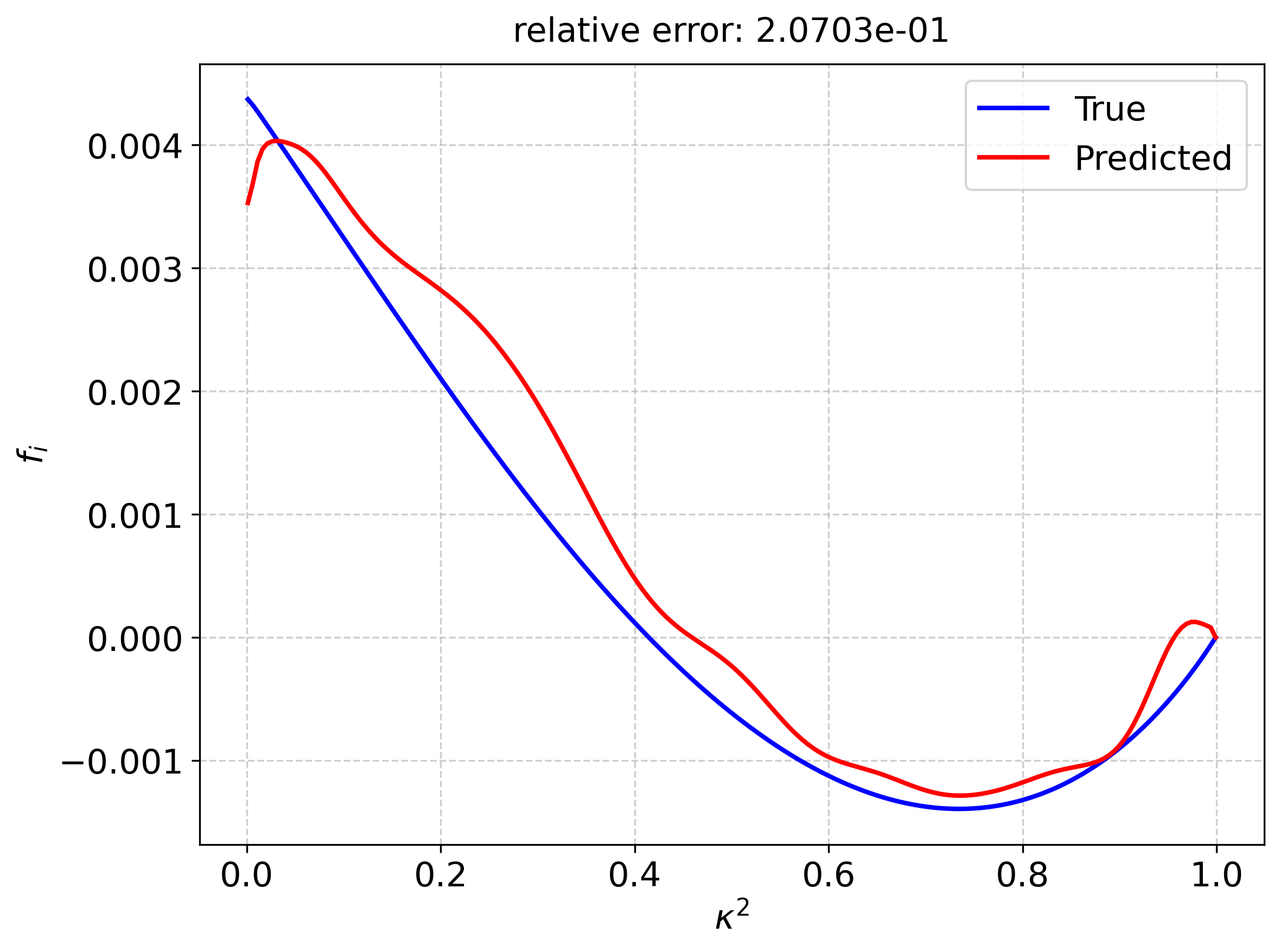}
	\put(-50,40){\textbf{\footnotesize{(b) $DKE_{data}$}}}
	\vfill
	\includegraphics[width=0.22\textwidth]{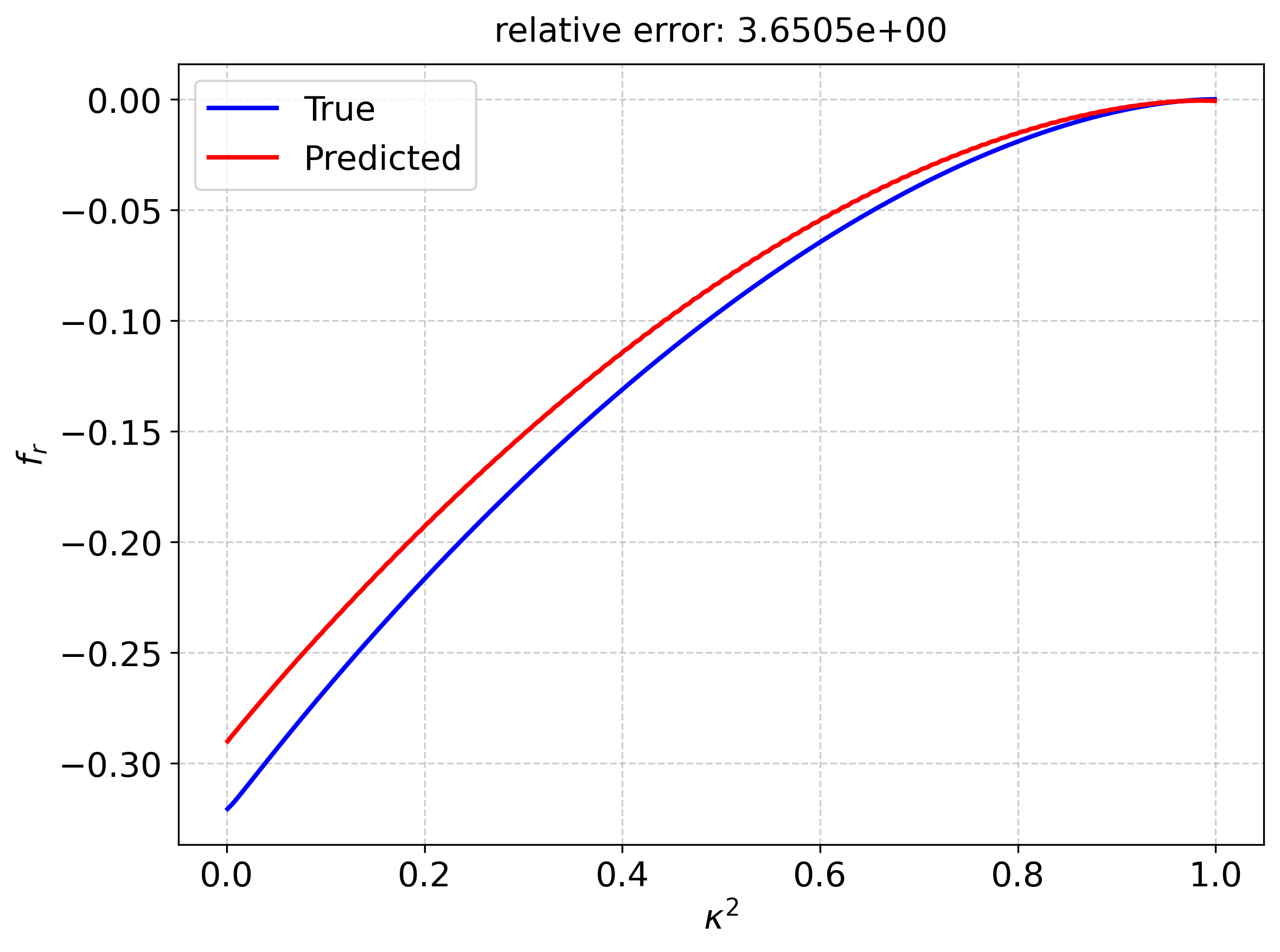}
	\put(-50,40){\textbf{\footnotesize{(c) $DKE_{phys}$}}}
	\includegraphics[width=0.22\textwidth]{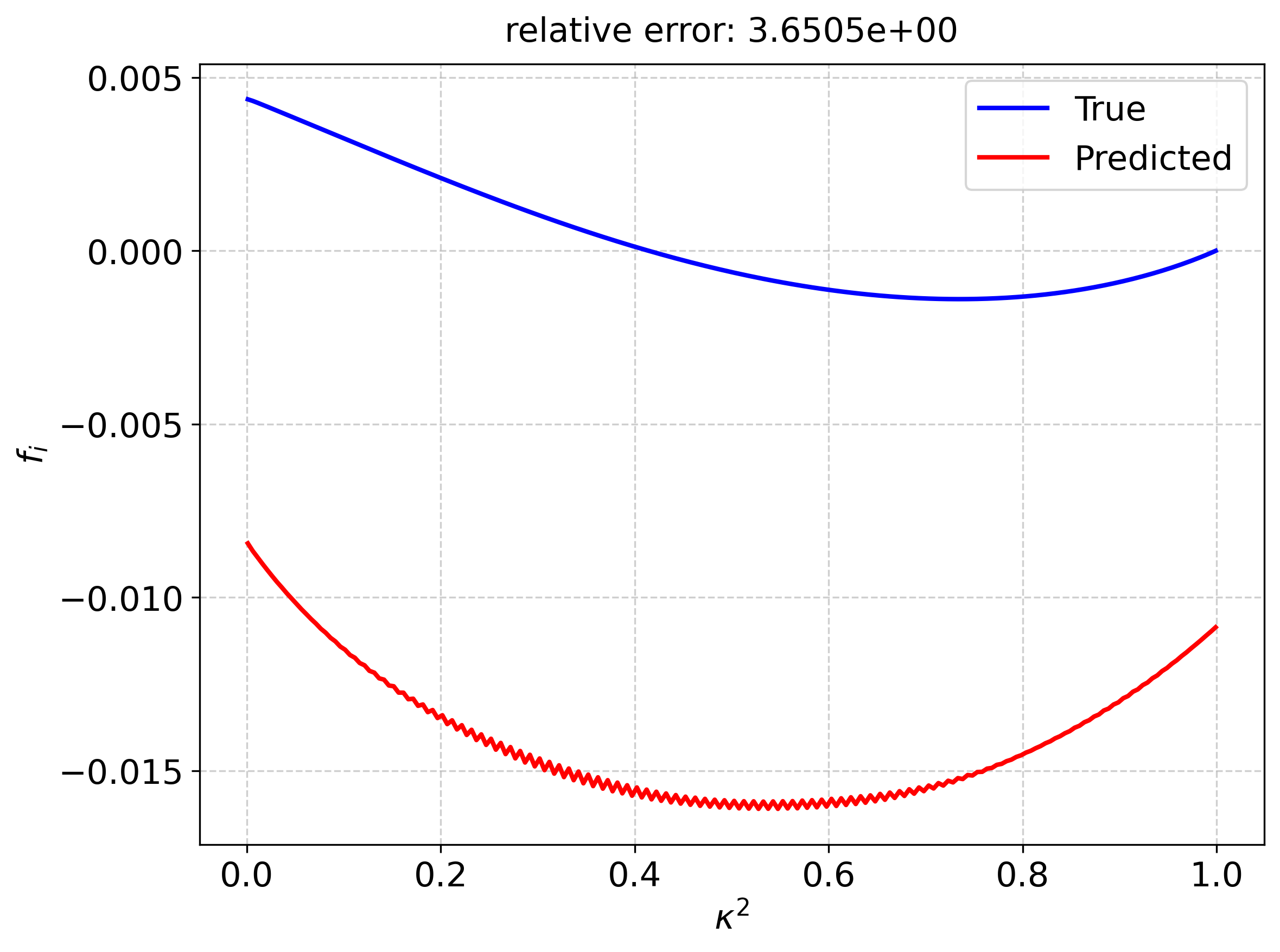}
	\put(-50,40){\textbf{\footnotesize{(d) $DKE_{phys}$}}}
	\vfill
	\includegraphics[width=0.22\textwidth]{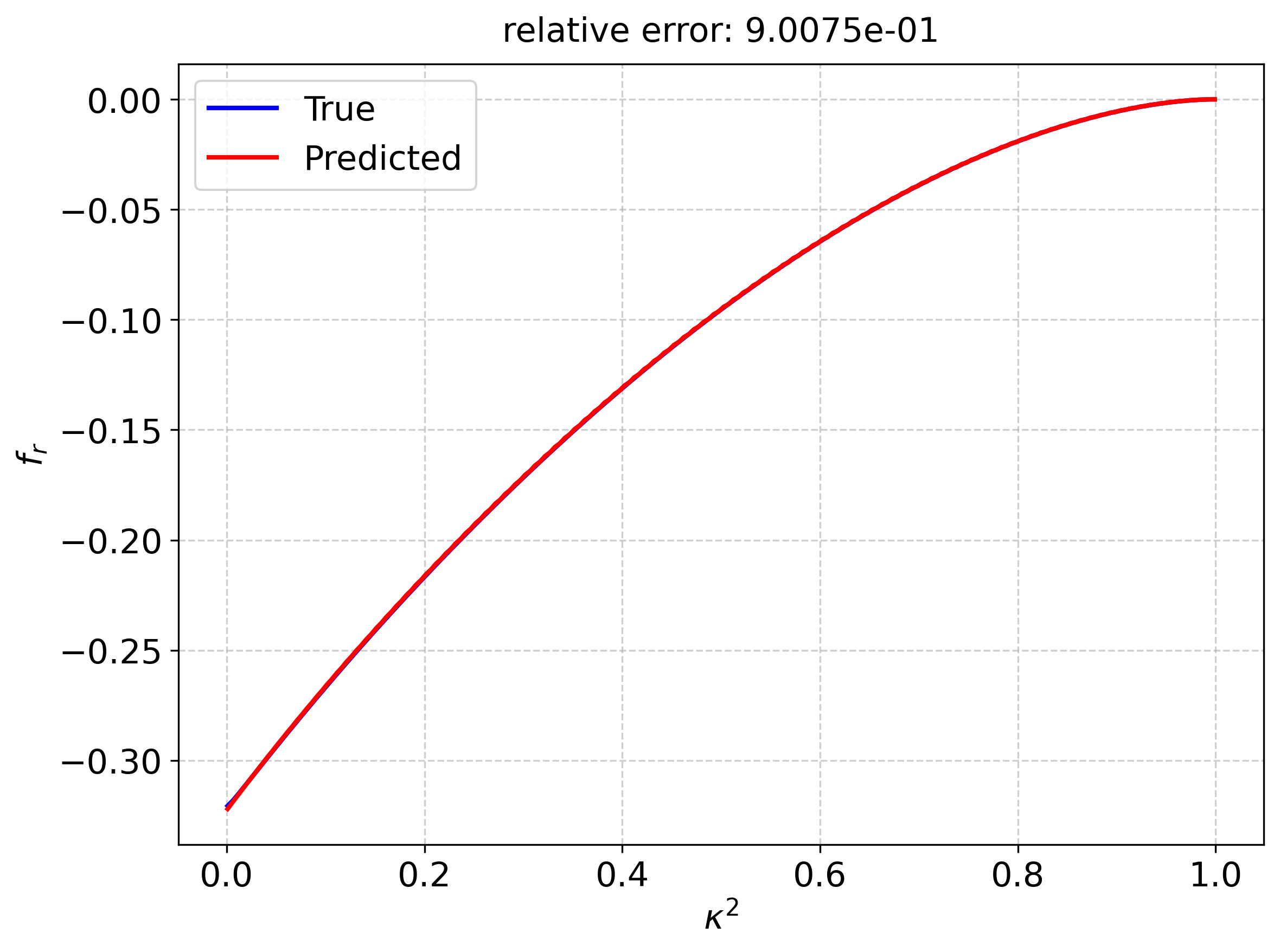}
	\put(-50,40){\textbf{\footnotesize{(e) $DKE_{phys,bc1}$}}}
	\includegraphics[width=0.22\textwidth]{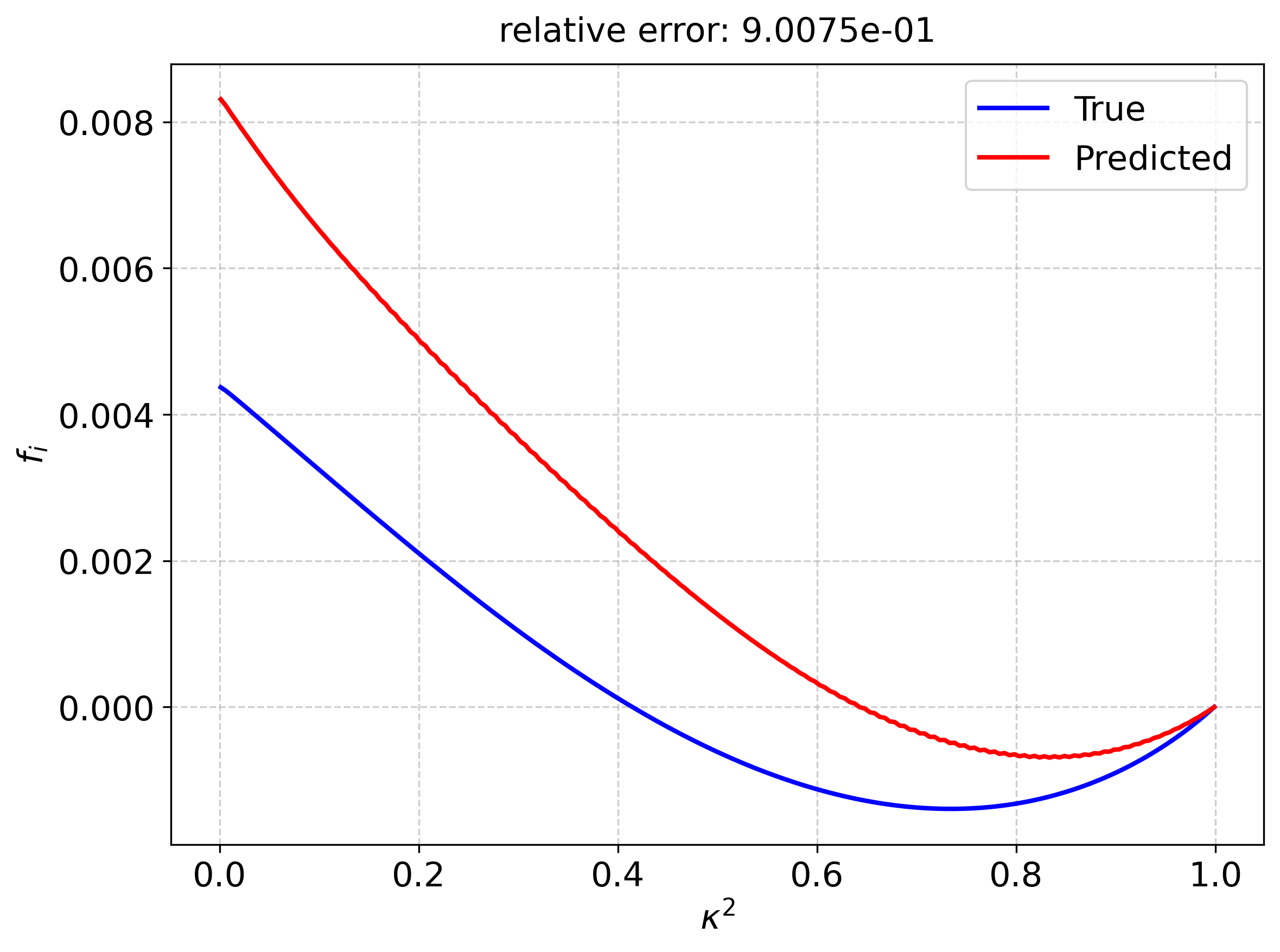}
	\put(-50,40){\textbf{\footnotesize{(f) $DKE_{phys,bc1}$}}}
	\caption{Prediction of DKE solution (left column for $f_r$ and right column for $f_i$) by surrogate models (top panel: $DKE_{data}$; middle panel: $DKE_{phys}$; bottom panel: $DKE_{phys,bc1}$) for one same data sample.}
	\label{fig:idx_sample}
\end{figure}

To further illustrate the difference of model performance among different approaches, Fig.~\ref{fig:idx_sample} shows prediction of three surrogates for one same data sample. Notably, the imaginary part predicted by $DKE_{data}$ (Fig.~\ref{fig:idx_sample}(b)) shows obvious unphysical bumpiness, although the relative error is the smallest among three surrogates. On the contrary, $DKE_{phys,bc1}$ correctly captures the overall shape of the DKE solution, although the relative error is large due to deviations around $\kappa^2\sim0$ (Fig.~\ref{fig:idx_sample}(f)). $DKE_{phys}$ surrogate also well captures the profile shape, but the large deviation at $\kappa^2=1$ boundary results in overall downward shift (Fig.~\ref{fig:idx_sample}(d)). Generally speaking, the data-driven approach usually exhibits high accuracy regarding data proximity, but unphysical predictions (e.g. profile bumpiness) may appear because no physical law is constrained. Physics-constrained approach shows generally better physical consistency, and the appropriate hard-constraint (e.g. boundary condition) plays significant roles in improving model performance.

\subsection{Prediction time}

\begin{figure}[htbp]
	\centering
	\includegraphics[width=0.48\textwidth]{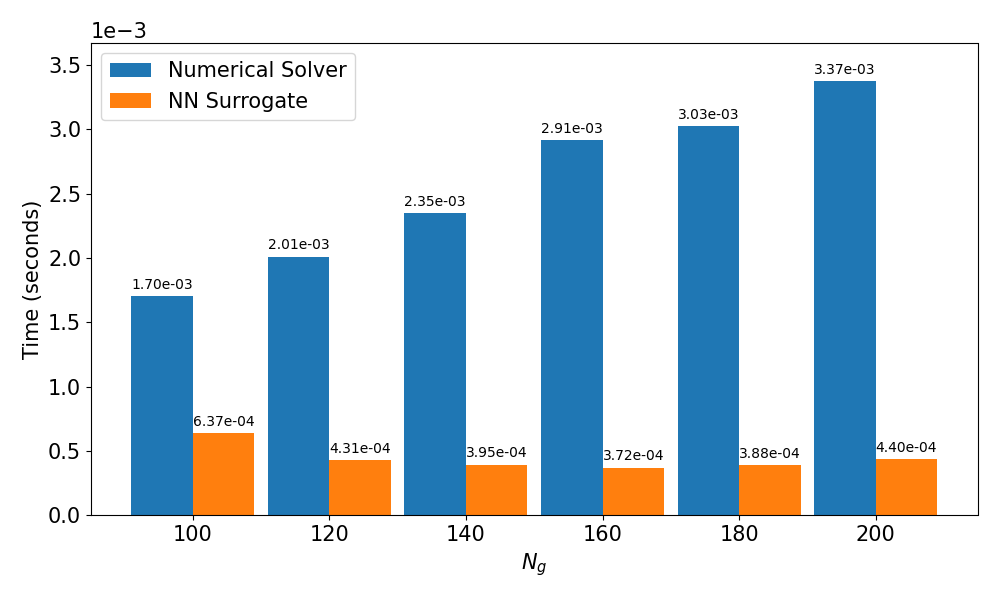}
	\caption{Comparison of calculation time between numerical DKE solver and deep learning surrogate for different $N_g$.}
	\label{fig:time}
\end{figure}

Fig.~\ref{fig:time} compares time consumption of forward calculation of surrogate solver and numerical DKE solver modeling, and their dependence on number of grid points $N_g$ (which directly influences size of input and output layers). Because three DKE surrogates share the same model architecture and hyper-parameters, their forward computational time is similar. The time comparison is performed on CPU only for fair comparison. The numerical solver execution time is the averaged time cost over $13130$ calls (which is the typical number of calls for one complete NTV calculation cycle), the NN surrogate calculation time is the average of $1000$ executions. For $N_g=200$, DKE surrogate accelerates DKE solving task by nearly one order of magnitude ($\times7.66$ speedup), thus facilitating further fast modeling of NTV torque. Time cost of surrogate solver is not a monotonic function of $N_g$, but the overall variation is relatively insignificant. This indicates that the input/output layer size is not the key influencing factor on time cost, the hidden layer calculation and the memory cost may also play important roles. However, with the increase of $N_g$, the numerical solver time cost increases monotonically, so as the computational speedup ratio. In other words, the acceleration of surrogate solver may become more significant in high-precision (i.e. large $N_g$) scientific computing tasks.

\section{Summary and future work}
\label{sec:sum}

The key ideas and results are illustrated in Fig.~\ref{fig:flowchart}. In summary, this work proposes a data-free, physics-constrained approach for developing fast and high-fidelity DKE surrogate solver, by utilizing the governing physical equations as loss function and implementing one of the boundary conditions as hard constraint. The physics-constrained surrogate solver exhibits similar prediction accuracy with data-driven approach regarding data proximity, while showing improved physical consistency indicated by lower physical loss amplitude and reduced profile bumpiness. In the meantime, the surrogate solver accelerates the traditional numerical DKE solver by nearly one order of magnitude, thus enabling further fast NTV torque modeling tasks. The physics-driven approach is proven to be a promising method to train fast surrogate for time-consuming scientific computing tasks, under data-scarce or data-free scenarios.

The future work will focus on the following aspects: First, the prediction accuracy of the physics-constrained surrogates can still be improved, for example by optimizing the model architecture, fine-tuning hyper-parameters, implementing adaptive weighting strategy and so on; Second, the boundary condition at $\kappa^2=0$ also needs to be handled carefully by hard-constrained approach, which can further reduce the degree of freedom of parameter space and accelerate model convergence; Finally, the surrogate solver will be integrated with NTV modeling framework to illustrate its capability of fast and high-fidelity NTV torque modeling, thus enabling its application in practical physics studies.

\section{GenAI Disclosure}

The authors used DeepSeek for language polishing and grammatical correction of the manuscript. The authors have reviewed all content generated by AI tool in detail, and take full responsibility for the entire manuscript. No GenAI tools were used in producing scientific results.

\begin{acks}
This work is supported by the National Key R\&D Program of China under Grant No. 2024YFE03010000.
\end{acks}

\bibliographystyle{ACM-Reference-Format}
\bibliography{refs}


\begin{thebibliography}{21}


\ifx \showCODEN    \undefined \def \showCODEN     #1{\unskip}     \fi
\ifx \showISBNx    \undefined \def \showISBNx     #1{\unskip}     \fi
\ifx \showISBNxiii \undefined \def \showISBNxiii  #1{\unskip}     \fi
\ifx \showISSN     \undefined \def \showISSN      #1{\unskip}     \fi
\ifx \showLCCN     \undefined \def \showLCCN      #1{\unskip}     \fi
\ifx \shownote     \undefined \def \shownote      #1{#1}          \fi
\ifx \showarticletitle \undefined \def \showarticletitle #1{#1}   \fi
\ifx \showURL      \undefined \def \showURL       {\relax}        \fi
\providecommand\bibfield[2]{#2}
\providecommand\bibinfo[2]{#2}
\providecommand\natexlab[1]{#1}
\providecommand\showeprint[2][]{arXiv:#2}

\bibitem[Bécoulet et~al\mbox{.}(2017)]%
        {BecouletM_2017_NF}
\bibfield{author}{\bibinfo{person}{M Bécoulet}, \bibinfo{person}{M Kim},
  \bibinfo{person}{G Yun}, \bibinfo{person}{S Pamela}, \bibinfo{person}{J
  Morales}, \bibinfo{person}{X Garbet}, \bibinfo{person}{G~T~A Huijsmans},
  \bibinfo{person}{C Passeron}, \bibinfo{person}{O Février},
  \bibinfo{person}{M Hoelzl}, \bibinfo{person}{A Lessig}, {and}
  \bibinfo{person}{F Orain}.} \bibinfo{year}{2017}\natexlab{}.
\newblock \showarticletitle{Non-linear {MHD} modelling of edge localized modes
  dynamics in {KSTAR}}.
\newblock \bibinfo{journal}{\emph{Nuclear Fusion}} \bibinfo{volume}{57},
  \bibinfo{number}{11} (\bibinfo{date}{Aug.} \bibinfo{year}{2017}),
  \bibinfo{pages}{116059}.
\newblock


\bibitem[Cho et~al\mbox{.}(2024)]%
        {ChoW_2024_ICML}
\bibfield{author}{\bibinfo{person}{Woojin Cho}, \bibinfo{person}{Minju Jo},
  \bibinfo{person}{Haksoo Lim}, \bibinfo{person}{Kookjin Lee},
  \bibinfo{person}{Dongeun Lee}, \bibinfo{person}{Sanghyun Hong}, {and}
  \bibinfo{person}{Noseong Park}.} \bibinfo{year}{2024}\natexlab{}.
\newblock \showarticletitle{Parameterized physics-informed neural networks for
  parameterized {PDEs}}. In \bibinfo{booktitle}{\emph{Proceedings of the 41st
  {International} {Conference} on {Machine} {Learning}}}
  \emph{(\bibinfo{series}{{ICML}'24})}. \bibinfo{publisher}{JMLR.org}.
\newblock
\newblock
\shownote{Place: Vienna, Austria}.


\bibitem[Clement et~al\mbox{.}(2021)]%
        {Clement_2021_NF}
\bibfield{author}{\bibinfo{person}{M~D Clement}, \bibinfo{person}{N~C Logan},
  {and} \bibinfo{person}{M~D Boyer}.} \bibinfo{year}{2021}\natexlab{}.
\newblock \showarticletitle{Neoclassical toroidal viscosity torque prediction
  via deep learning}.
\newblock \bibinfo{journal}{\emph{Nuclear Fusion}} \bibinfo{volume}{62},
  \bibinfo{number}{2} (\bibinfo{date}{Dec.} \bibinfo{year}{2021}),
  \bibinfo{pages}{26022}.
\newblock


\bibitem[Cuomo et~al\mbox{.}(2022)]%
        {CuomoS_2022_JSC}
\bibfield{author}{\bibinfo{person}{Salvatore Cuomo},
  \bibinfo{person}{Vincenzo~Schiano Di~Cola}, \bibinfo{person}{Fabio
  Giampaolo}, \bibinfo{person}{Gianluigi Rozza}, \bibinfo{person}{Maziar
  Raissi}, {and} \bibinfo{person}{Francesco Piccialli}.}
  \bibinfo{year}{2022}\natexlab{}.
\newblock \showarticletitle{Scientific {Machine} {Learning} {Through}
  {Physics}–{Informed} {Neural} {Networks}: {Where} we are and {What}’s
  {Next}}.
\newblock \bibinfo{journal}{\emph{Journal of Scientific Computing}}
  \bibinfo{volume}{92}, \bibinfo{number}{3} (\bibinfo{date}{July}
  \bibinfo{year}{2022}), \bibinfo{pages}{88}.
\newblock
\showISSN{1573-7691}


\bibitem[He et~al\mbox{.}(2016)]%
        {HeK_2016_CVPR}
\bibfield{author}{\bibinfo{person}{Kaiming He}, \bibinfo{person}{Xiangyu
  Zhang}, \bibinfo{person}{Shaoqing Ren}, {and} \bibinfo{person}{Jian Sun}.}
  \bibinfo{year}{2016}\natexlab{}.
\newblock \showarticletitle{Deep {Residual} {Learning} for {Image}
  {Recognition}}. In \bibinfo{booktitle}{\emph{2016 {IEEE} {Conference} on
  {Computer} {Vision} and {Pattern} {Recognition} ({CVPR})}}.
  \bibinfo{pages}{770--778}.
\newblock


\bibitem[Karniadakis et~al\mbox{.}(2021)]%
        {KarniadakisG_2021_NRP}
\bibfield{author}{\bibinfo{person}{George~Em Karniadakis},
  \bibinfo{person}{Ioannis~G. Kevrekidis}, \bibinfo{person}{Lu Lu},
  \bibinfo{person}{Paris Perdikaris}, \bibinfo{person}{Sifan Wang}, {and}
  \bibinfo{person}{Liu Yang}.} \bibinfo{year}{2021}\natexlab{}.
\newblock \showarticletitle{Physics-informed machine learning}.
\newblock \bibinfo{journal}{\emph{Nature Reviews Physics}} \bibinfo{volume}{3},
  \bibinfo{number}{6} (\bibinfo{date}{June} \bibinfo{year}{2021}),
  \bibinfo{pages}{422--440}.
\newblock
\showISSN{2522-5820}


\bibitem[Kim et~al\mbox{.}(2024)]%
        {KimS_2024_NC}
\bibfield{author}{\bibinfo{person}{S.~K. Kim}, \bibinfo{person}{R. Shousha},
  \bibinfo{person}{S.~M. Yang}, \bibinfo{person}{Q. Hu}, \bibinfo{person}{S.~H.
  Hahn}, \bibinfo{person}{A. Jalalvand}, \bibinfo{person}{J.-K. Park},
  \bibinfo{person}{N.~C. Logan}, \bibinfo{person}{A.~O. Nelson},
  \bibinfo{person}{Y.-S. Na}, \bibinfo{person}{R. Nazikian},
  \bibinfo{person}{R. Wilcox}, \bibinfo{person}{R. Hong}, \bibinfo{person}{T.
  Rhodes}, \bibinfo{person}{C. Paz-Soldan}, \bibinfo{person}{Y.~M. Jeon},
  \bibinfo{person}{M.~W. Kim}, \bibinfo{person}{W.~H. Ko},
  \bibinfo{person}{J.~H. Lee}, \bibinfo{person}{A. Battey}, \bibinfo{person}{G.
  Yu}, \bibinfo{person}{A. Bortolon}, \bibinfo{person}{J. Snipes}, {and}
  \bibinfo{person}{E. Kolemen}.} \bibinfo{year}{2024}\natexlab{}.
\newblock \showarticletitle{Highest fusion performance without harmful edge
  energy bursts in tokamak}.
\newblock \bibinfo{journal}{\emph{Nature Communications}} \bibinfo{volume}{15},
  \bibinfo{number}{1} (\bibinfo{date}{May} \bibinfo{year}{2024}),
  \bibinfo{pages}{3990}.
\newblock
\showISSN{2041-1723}


\bibitem[Loarte et~al\mbox{.}(2025)]%
        {LoarteA_2025_NF}
\bibfield{author}{\bibinfo{person}{A Loarte}, \bibinfo{person}{R~A Pitts},
  \bibinfo{person}{T Wauters}, \bibinfo{person}{I Nunes}, \bibinfo{person}{P de
  Vries}, \bibinfo{person}{S~H Kim}, \bibinfo{person}{F Köchl},
  \bibinfo{person}{A Polevoi}, \bibinfo{person}{M Lehnen}, \bibinfo{person}{J
  Artola}, \bibinfo{person}{S Jachmich}, \bibinfo{person}{A Pshenov},
  \bibinfo{person}{X Bai}, \bibinfo{person}{I~S Carvalho}, \bibinfo{person}{M
  Dubrov}, \bibinfo{person}{Y Gribov}, \bibinfo{person}{M Schneider},
  \bibinfo{person}{L Zabeo}, \bibinfo{person}{X Bonnin}, \bibinfo{person}{S~D
  Pinches}, \bibinfo{person}{F Poli}, \bibinfo{person}{G~Suarez Lopez},
  \bibinfo{person}{M Merola}, \bibinfo{person}{F Escourbiac},
  \bibinfo{person}{R Hunt}, \bibinfo{person}{L Chen}, \bibinfo{person}{D
  Boilson}, \bibinfo{person}{P Veltri}, \bibinfo{person}{N Casal},
  \bibinfo{person}{M Preynas}, \bibinfo{person}{A Mukherjee},
  \bibinfo{person}{W Helou}, \bibinfo{person}{F Kazarian}, \bibinfo{person}{S
  Willms}, \bibinfo{person}{I Bonnet}, \bibinfo{person}{R Michling},
  \bibinfo{person}{L Giancarli}, \bibinfo{person}{J van~der Laan},
  \bibinfo{person}{M Walsh}, \bibinfo{person}{V Udintsev}, \bibinfo{person}{R
  Reichle}, \bibinfo{person}{G Vayakis}, \bibinfo{person}{A Fossen},
  \bibinfo{person}{M Turnyanskiy}, \bibinfo{person}{A Becoulet},
  \bibinfo{person}{Y Kamada}, \bibinfo{person}{G Zhuang}, \bibinfo{person}{G
  Xu}, \bibinfo{person}{X Gong}, \bibinfo{person}{J Huang}, \bibinfo{person}{M
  Jia}, \bibinfo{person}{R Ding}, \bibinfo{person}{J Qian}, \bibinfo{person}{Y
  Sun}, \bibinfo{person}{Q Yang}, \bibinfo{person}{L Zhang}, \bibinfo{person}{M
  Xu}, \bibinfo{person}{L Zhang}, \bibinfo{person}{S Brezinsek},
  \bibinfo{person}{J Stober}, \bibinfo{person}{J Hobirk}, \bibinfo{person}{F
  Rimini}, \bibinfo{person}{J Garcia}, \bibinfo{person}{S~L Rao},
  \bibinfo{person}{J Ghosh}, \bibinfo{person}{D Sharma}, \bibinfo{person}{B
  Magesh}, \bibinfo{person}{R~P Bhattacharya}, \bibinfo{person}{G Matsunaga},
  \bibinfo{person}{H Urano}, \bibinfo{person}{T Hirose}, \bibinfo{person}{K
  Ogawa}, \bibinfo{person}{G Motojima}, \bibinfo{person}{C~K Sung},
  \bibinfo{person}{H~H Lee}, \bibinfo{person}{J~K Park}, \bibinfo{person}{M~S
  Cheon}, \bibinfo{person}{Y~M Jeon}, \bibinfo{person}{S Konovalov},
  \bibinfo{person}{S Lebedev}, \bibinfo{person}{N Kirneva}, \bibinfo{person}{Y
  Kashchuk}, \bibinfo{person}{N Bakharev}, \bibinfo{person}{X Chen},
  \bibinfo{person}{A Bortolon}, \bibinfo{person}{L Casali}, \bibinfo{person}{R
  Maingi}, \bibinfo{person}{F Turco}, \bibinfo{person}{K Schmid},
  \bibinfo{person}{Y Liu}, \bibinfo{person}{J~R Martín-Solís},
  \bibinfo{person}{C Angioni}, \bibinfo{person}{I Pusztai}, \bibinfo{person}{D
  Fajardo}, \bibinfo{person}{D Mateev}, \bibinfo{person}{E Lerche},
  \bibinfo{person}{D van Eester}, \bibinfo{person}{P Vincenzi},
  \bibinfo{person}{R Futtersack}, \bibinfo{person}{V Bobkov}, {and}
  \bibinfo{person}{L Colas}.} \bibinfo{year}{2025}\natexlab{}.
\newblock \showarticletitle{The new {ITER} baseline, research plan and open
  {R}\&amp;{D} issues}.
\newblock \bibinfo{journal}{\emph{Plasma Physics and Controlled Fusion}}
  \bibinfo{volume}{67}, \bibinfo{number}{6} (\bibinfo{date}{June}
  \bibinfo{year}{2025}), \bibinfo{pages}{065023}.
\newblock


\bibitem[Logan et~al\mbox{.}(2022)]%
        {LoganN_2022_PRL}
\bibfield{author}{\bibinfo{person}{N.~C. Logan}, \bibinfo{person}{Q. Hu},
  \bibinfo{person}{C. Paz-Soldan}, \bibinfo{person}{R. Nazikian},
  \bibinfo{person}{T. Rhodes}, \bibinfo{person}{T. Wilks}, \bibinfo{person}{S.
  Munaretto}, \bibinfo{person}{A. Bortolon}, \bibinfo{person}{F. Laggner},
  \bibinfo{person}{F. Scotti}, \bibinfo{person}{R. Hong}, {and}
  \bibinfo{person}{H. Wang}.} \bibinfo{year}{2022}\natexlab{}.
\newblock \showarticletitle{Improved {Particle} {Confinement} with {Resonant}
  {Magnetic} {Perturbations} in {DIII}-{D} {Tokamak} {H}-{Mode} {Plasmas}}.
\newblock \bibinfo{journal}{\emph{Phys. Rev. Lett.}} \bibinfo{volume}{129},
  \bibinfo{number}{20} (\bibinfo{date}{Nov.} \bibinfo{year}{2022}),
  \bibinfo{pages}{205001}.
\newblock


\bibitem[Logan et~al\mbox{.}(2021)]%
        {LoganN_2021_NF}
\bibfield{author}{\bibinfo{person}{N~C Logan}, \bibinfo{person}{C Zhu},
  \bibinfo{person}{J.-K. Park}, \bibinfo{person}{S~M Yang}, {and}
  \bibinfo{person}{Q Hu}.} \bibinfo{year}{2021}\natexlab{}.
\newblock \showarticletitle{Physics basis for design of {3D} coils in
  tokamaks}.
\newblock \bibinfo{journal}{\emph{Nuclear Fusion}} \bibinfo{volume}{61},
  \bibinfo{number}{7} (\bibinfo{date}{June} \bibinfo{year}{2021}),
  \bibinfo{pages}{76010}.
\newblock


\bibitem[Park et~al\mbox{.}(2009)]%
        {Park_2009_PRL}
\bibfield{author}{\bibinfo{person}{Jong-Kyu Park}, \bibinfo{person}{Allen~H
  Boozer}, {and} \bibinfo{person}{Jonathan~E Menard}.}
  \bibinfo{year}{2009}\natexlab{}.
\newblock \showarticletitle{Nonambipolar Transport by Trapped Particles in
  Tokamaks}.
\newblock \bibinfo{journal}{\emph{Physical Review Letters}}
  \bibinfo{volume}{102} (\bibinfo{date}{2} \bibinfo{year}{2009}),
  \bibinfo{pages}{65002}.
\newblock
Issue 6.


\bibitem[Raissi et~al\mbox{.}(2019)]%
        {RaissiM_2019_JCP}
\bibfield{author}{\bibinfo{person}{M. Raissi}, \bibinfo{person}{P. Perdikaris},
  {and} \bibinfo{person}{G.~E. Karniadakis}.} \bibinfo{year}{2019}\natexlab{}.
\newblock \showarticletitle{Physics-informed neural networks: {A} deep learning
  framework for solving forward and inverse problems involving nonlinear
  partial differential equations}.
\newblock \bibinfo{journal}{\emph{J. Comput. Phys.}}  \bibinfo{volume}{378}
  (\bibinfo{year}{2019}), \bibinfo{pages}{686--707}.
\newblock
\showISSN{0021-9991}


\bibitem[Shaing(2003)]%
        {Shaing_2003_POP}
\bibfield{author}{\bibinfo{person}{K~C Shaing}.}
  \bibinfo{year}{2003}\natexlab{}.
\newblock \showarticletitle{Magnetohydrodynamic-activity-induced toroidal
  momentum dissipation in collisionless regimes in tokamaks}.
\newblock \bibinfo{journal}{\emph{Physics of Plasmas}}  \bibinfo{volume}{10}
  (\bibinfo{year}{2003}), \bibinfo{pages}{1443--1448}.
\newblock
Issue 5.


\bibitem[Shaing et~al\mbox{.}(2015)]%
        {Shaing_2015_NF}
\bibfield{author}{\bibinfo{person}{K~C Shaing}, \bibinfo{person}{K Ida}, {and}
  \bibinfo{person}{S~A Sabbagh}.} \bibinfo{year}{2015}\natexlab{}.
\newblock \showarticletitle{Neoclassical plasma viscosity and transport
  processes in non-axisymmetric tori}.
\newblock \bibinfo{journal}{\emph{Nuclear Fusion}}  \bibinfo{volume}{55}
  (\bibinfo{date}{11} \bibinfo{year}{2015}), \bibinfo{pages}{125001}.
\newblock
Issue 12.


\bibitem[Sun et~al\mbox{.}(2019)]%
        {Sun_2019_POP}
\bibfield{author}{\bibinfo{person}{Y Sun}, \bibinfo{person}{X Li},
  \bibinfo{person}{K He}, {and} \bibinfo{person}{K~C Shaing}.}
  \bibinfo{year}{2019}\natexlab{}.
\newblock \showarticletitle{Unified modeling of both resonant and non-resonant
  neoclassical transport under non-axisymmetric magnetic perturbations in
  tokamaks}.
\newblock \bibinfo{journal}{\emph{Physics of Plasmas}}  \bibinfo{volume}{26}
  (\bibinfo{year}{2019}), \bibinfo{pages}{72504}.
\newblock
Issue 7.


\bibitem[Sun et~al\mbox{.}(2011)]%
        {Sun_2011_NF}
\bibfield{author}{\bibinfo{person}{Y. Sun}, \bibinfo{person}{Y. Liang},
  \bibinfo{person}{K.C. Shaing}, \bibinfo{person}{H.R. Koslowski},
  \bibinfo{person}{C. Wiegmann}, {and} \bibinfo{person}{T. Zhang}.}
  \bibinfo{year}{2011}\natexlab{}.
\newblock \showarticletitle{Modelling of the neoclassical toroidal plasma
  viscosity torque in tokamaks}.
\newblock \bibinfo{journal}{\emph{Nuclear Fusion}} \bibinfo{volume}{51},
  \bibinfo{number}{5} (\bibinfo{date}{April} \bibinfo{year}{2011}),
  \bibinfo{pages}{053015}.
\newblock


\bibitem[Sun et~al\mbox{.}(2010)]%
        {Sun_2010_PRL}
\bibfield{author}{\bibinfo{person}{Y. Sun}, \bibinfo{person}{Y. Liang},
  \bibinfo{person}{K.~C. Shaing}, \bibinfo{person}{H.~R. Koslowski},
  \bibinfo{person}{C. Wiegmann}, {and} \bibinfo{person}{T. Zhang}.}
  \bibinfo{year}{2010}\natexlab{}.
\newblock \showarticletitle{Neoclassical {Toroidal} {Plasma} {Viscosity}
  {Torque} in {Collisionless} {Regimes} in {Tokamaks}}.
\newblock \bibinfo{journal}{\emph{Physical Review Letters}}
  \bibinfo{volume}{105}, \bibinfo{number}{14} (\bibinfo{date}{Oct.}
  \bibinfo{year}{2010}), \bibinfo{pages}{145002}.
\newblock


\bibitem[Wan et~al\mbox{.}(2017)]%
        {WanB_2017_NF}
\bibfield{author}{\bibinfo{person}{B~N Wan}, \bibinfo{person}{Y~F Liang},
  \bibinfo{person}{X~Z Gong}, \bibinfo{person}{J~G Li}, \bibinfo{person}{N
  Xiang}, \bibinfo{person}{G~S Xu}, \bibinfo{person}{Y~W Sun},
  \bibinfo{person}{L Wang}, \bibinfo{person}{J~P Qian}, \bibinfo{person}{H~Q
  Liu}, \bibinfo{person}{X~D Zhang}, \bibinfo{person}{L~Q Hu},
  \bibinfo{person}{J~S Hu}, \bibinfo{person}{F~K Liu}, \bibinfo{person}{C~D
  Hu}, \bibinfo{person}{Y~P Zhao}, \bibinfo{person}{L Zeng}, \bibinfo{person}{M
  Wang}, \bibinfo{person}{H~D Xu}, \bibinfo{person}{G~N Luo},
  \bibinfo{person}{A~M Garofalo}, \bibinfo{person}{A Ekedahl},
  \bibinfo{person}{L Zhang}, \bibinfo{person}{X~J Zhang}, \bibinfo{person}{J
  Huang}, \bibinfo{person}{B~J Ding}, \bibinfo{person}{Q Zang},
  \bibinfo{person}{M~H Li}, \bibinfo{person}{F Ding}, \bibinfo{person}{S~Y
  Ding}, \bibinfo{person}{B Lyu}, \bibinfo{person}{Y~W Yu}, \bibinfo{person}{T
  Zhang}, \bibinfo{person}{Y Zhang}, \bibinfo{person}{G~Q Li},
  \bibinfo{person}{T~Y Xia}, \bibinfo{person}{{the EAST team}}, {and}
  \bibinfo{person}{{Collaborators}}.} \bibinfo{year}{2017}\natexlab{}.
\newblock \showarticletitle{Overview of {EAST} experiments on the development
  of high-performance steady-state scenario}.
\newblock \bibinfo{journal}{\emph{Nuclear Fusion}} \bibinfo{volume}{57},
  \bibinfo{number}{10} (\bibinfo{date}{July} \bibinfo{year}{2017}),
  \bibinfo{pages}{102019}.
\newblock


\bibitem[Wu et~al\mbox{.}(2025)]%
        {WuX_2025_NF}
\bibfield{author}{\bibinfo{person}{X.M. Wu}, \bibinfo{person}{Y. Sun},
  \bibinfo{person}{Q. Ma}, \bibinfo{person}{S. Gu}, \bibinfo{person}{Y.F.
  Wang}, \bibinfo{person}{K.N. Geng}, \bibinfo{person}{X.X. Zhang},
  \bibinfo{person}{G.Q. Li}, \bibinfo{person}{H. Sheng}, \bibinfo{person}{T.
  Zhang}, \bibinfo{person}{P.B. Snyder}, \bibinfo{person}{Q. Zhang},
  \bibinfo{person}{J. qian}, \bibinfo{person}{Y.Y. Li}, {and}
  \bibinfo{person}{B. Wan}.} \bibinfo{year}{2025}\natexlab{}.
\newblock \showarticletitle{Influence of n = 4 {RMPs} on pedestal structure and
  stability in {EAST}}.
\newblock \bibinfo{journal}{\emph{Nuclear Fusion}} \bibinfo{volume}{65},
  \bibinfo{number}{7} (\bibinfo{date}{June} \bibinfo{year}{2025}),
  \bibinfo{pages}{076031}.
\newblock


\bibitem[Yan et~al\mbox{.}(2025)]%
        {Yan_2025_CPC}
\bibfield{author}{\bibinfo{person}{X.-T. Yan}, \bibinfo{person}{N.-N. Bao},
  \bibinfo{person}{C.-Y. Zhao}, \bibinfo{person}{Y.-W. Sun},
  \bibinfo{person}{Y.-T. Meng}, \bibinfo{person}{W.-Y. Zhou},
  \bibinfo{person}{N.-Y. Liang}, \bibinfo{person}{Y.-X. Lu},
  \bibinfo{person}{Y.-F. Liang}, {and} \bibinfo{person}{B.-N. Wan}.}
  \bibinfo{year}{2025}\natexlab{}.
\newblock \showarticletitle{{NTVTOK}-{ML}: {Fast} surrogate model for
  neoclassical toroidal viscosity torque calculation in tokamaks based on
  machine learning methods}.
\newblock \bibinfo{journal}{\emph{Computer Physics Communications}}
  \bibinfo{volume}{307} (\bibinfo{date}{Feb.} \bibinfo{year}{2025}),
  \bibinfo{pages}{109413}.
\newblock


\bibitem[Yang et~al\mbox{.}(2019)]%
        {Yang_2019_PRL}
\bibfield{author}{\bibinfo{person}{S~M Yang}, \bibinfo{person}{J.-K. Park},
  \bibinfo{person}{Yong-Su Na}, \bibinfo{person}{Z~R Wang},
  \bibinfo{person}{W~H Ko}, \bibinfo{person}{Y In}, \bibinfo{person}{J~H Lee},
  \bibinfo{person}{K~D Lee}, {and} \bibinfo{person}{S~K Kim}.}
  \bibinfo{year}{2019}\natexlab{}.
\newblock \showarticletitle{Nonambipolar Transport due to Electrons with 3D
  Resistive Response in the KSTAR Tokamak}.
\newblock \bibinfo{journal}{\emph{Physical Review Letters}}
  \bibinfo{volume}{123} (\bibinfo{date}{8} \bibinfo{year}{2019}),
  \bibinfo{pages}{95001}.
\newblock
Issue 9.


\end{thebibliography}


\end{document}